\documentclass[a4paper, 11pt]{article}
\usepackage{srcltx}
\usepackage[colorlinks=false,bookmarks=true]{hyperref}
\usepackage{amsfonts,amsmath,amsthm,amssymb,latexsym,enumerate,subfigure,appendix}
\usepackage[latin2]{inputenc}
\usepackage[english]{babel}
\usepackage{enumerate}
\usepackage{url}
\usepackage{lastpage}
\usepackage{longtable}
\usepackage{fancyvrb}
\usepackage{graphics}
\usepackage{pgf}
\usepackage{subfigure}
\usepackage[center]{caption}

\newcommand{\unitInt}{$[0,1]$}
\usepackage{authblk}

\begin{document}
\title{\textbf{Extracting information from the signature of a financial data stream}}
\author[$\dag$]{Lajos Gergely Gyurk\'o}
\author[$\dag$]{Terry Lyons}
\author[$\ddag$]{Mark Kontkowski}
\author[$\ddag$]{Jonathan Field}
\affil[$\dag$]{Oxford-Man Institute of Quantitative Finance, University of
  Oxford}
\affil[$\ddag$]{Man Group Plc}

\maketitle

\begin{abstract}
Market events such as order placement and order cancellation are
examples of the complex and substantial flow of data that surrounds a
modern financial engineer. New mathematical techniques, developed to
describe the interactions of complex oscillatory systems (known as the
theory of rough paths) provides new tools for analysing and describing
these data streams and extracting the vital information. In this paper
we illustrate how a very small number of coefficients obtained from
the signature of financial data can be sufficient to classify this
data for subtle underlying features and make useful predictions. 

This paper presents financial examples in which we learn from data and then proceed to classify fresh streams. The classification is based on features of streams that are specified through the coordinates of the signature of the path. At a mathematical level the signature is a faithful transform of a multidimensional time series. (Ben Hambly and Terry Lyons \cite{uniqueSig}), Hao Ni and Terry Lyons \cite{NiLyons} introduced the possibility of its use to understand financial data and pointed to the potential this approach has for machine learning and prediction. 

We evaluate and refine these theoretical suggestions against practical examples of interest and present a few motivating experiments which demonstrate information the signature can easily capture in a non-parametric way avoiding traditional statistical modelling of the data. In the first experiment we identify atypical market behaviour across standard 30-minute time buckets sampled from the WTI crude oil future market (NYMEX). The second and third experiments aim to characterise the market "impact" of and distinguish between parent orders generated by two different trade execution algorithms on the FTSE 100 Index futures market listed on NYSE Liffe.
\end{abstract}

\section{Introduction}


The frequency of market events such as order placement and order
cancellation in most trading venues has increased dramatically
since the introduction of computer based trading. 
During certain time intervals, orders
can be submitted or modified within milliseconds or even faster. It is
highly challenging to understand and track the characteristics of a market
at a given moment. Quantitative analysts are eager to find methods
to compress high frequency data streams without losing information
that is relevant for their objectives. 

This paper presents financial examples in which we classify streams of
data. The classification is based on features of streams that are
specified by a property of sample paths known
as the \emph{signature} of the path. 

At an abstract mathematical level, the notion of a signature as an
informative transform of a multidimensional time series was
established by Ben Hambly and Terry Lyons \cite{uniqueSig},
meanwhile Hao Ni and Terry Lyons \cite{NiLyons}  have introduced the
possibility of its use to understand financial data and pointed to the
power this approach has for machine learning and prediction. 

We evaluate and refine these theoretical suggestions against some real
world data and practical examples. In each of the analysed cases, we
identify particular features, that is a
low dimensional part of the signature that captures 
information relevant for the characterisation of various
features.

In section \ref{sec:signatureMain}, we recall the definition and key
properties of the signature of a data stream. Moreover, we present a
few motivating  examples which demonstrate what
information the signature  preserves with special attention on properties which are
not captured by the traditional statistical indicators. 

Section \ref{sec:methodology} gives a description of the input data as
well as the linear regression-based classification method. 

Finally, in section \ref{sec:numericalResults}, three numerical
experiments are presented and the classification accuracy is
analysed. In the first experiment we characterise the typical market
behaviour within 
standard 30-minute time buckets sampled from the WTI crude oil future
market (NYMEX). The second and third experiments aim to characterise the
market impact of parent orders generated by two different trade execution
algorithms 
on the FTSE 100 Index futures market listed on NYSE Liffe.

\section{Signature of data streams}\label{sec:signatureMain}

\subsection{Origins of signature}
The mathematical properties of iterated integrals of
piece-wise regular multi-dimensional paths were first
studied by K.T. Chen \cite{Chen57,Chen58}. B. Hambly and
T. Lyons \cite{uniqueSig} extended these results to continuous paths of bounded
variation. 

Iterated integrals  of continuous multi-dimensional paths naturally
arise in the Taylor-expansion of controlled ordinary differential
equations (\cite{roughPaths98, roughPaths02, SaintFlour}). Moreover,
several numerical approximations of the solution to stochastic
differential equations (SDEs) are based on the iterated (stochastic)
integrals of Brownian motion (the reader is referred to
\cite{KloedenPlaten} and the references therein). 

Informally, the signature as data transform maps multi-dimensional paths to the
sequence of their iterated integrals \eqref{eq:iteratedIntegral}, where the sequence is equipped
with some algebraic operations (ref. \cite{roughPaths98,
  roughPaths02, SaintFlour}). The signatures of continuous paths
play a central role in the theory of rough paths \cite{roughPaths98,
  roughPaths02, SaintFlour,FrizVictoir}. Moreover, the algebraic properties of the
signature have been extensively used in the construction of numerical methods
 for approximation of solutions to SDEs \cite{Kusuoka04, cubature,
   RPNumSDE} as well as for the approximation of solutions to
 differential equations driven by rough paths \cite{FrizVictoir,gyurkoThesis}. 

The signature of time series and its application have been established
by H. Ni and T. Lyons \cite{NiLyons}. 

\subsection{Intuitive introduction}\label{sec:intuitiveIntroduction}
Let $X$ be a continuous function of finite length defined on $[0,T]$ and taking values
in $\mathbb{R}^d$, where
$X_t=(X^1_t,\dots,X^d_t)$. Moreover, let $\mathcal{I}$ denote the
set of all multi-indices $(i_1,\dots,i_k)$, where $k\ge 0$, $i_j\in\{1,\dots,d\}$
for $j=1,\dots,k$. We introduce the notation for $0\le s < t \le T$:
\begin{equation}
X^I_{s,t} = \int_{s<u_1<\cdots<u_k<t}
 \text{d}X^{i_1}_{u_1}\cdots\text{d} X^{i_k}_{u_k},
\label{eq:iteratedIntegral}
\end{equation}
where $I=(i_1,\dots,i_k)$ is a multi-index with
$i_j\in\{1,2,\dots,d\}$ for $j=1,\dots,k$. 

The \emph{signature} $S_{s,t}(X)$ of $X$ over the time interval $[s,t]$ maps $X$ to
the sequence $(X^I_{s,t})_{I\in\mathcal{I}}.$ 

The following properties of iterated integrals and the signature are
relevant in this paper\footnote{For the rigorous treatment of these
  properties the reader is referred to \cite{SaintFlour,uniqueSig}}. 
\begin{enumerate}[(i)]
\item\label{item:timeRep} \textbf{Invariance to time re-parameterisation}: for any continuous and monoton
  increasing function $\sigma:[0,T]\to[S,U]$,
  $(i_1,\dots,i_k)\in\mathcal{I}$ and $0\le s<t\le T$
\[
\int_{s<u_1<\cdots<u_k<t}
 \text{d}X^{i_1}_{u_1}\cdots\text{d} X^{i_k}_{u_k} = 
\int_{\sigma(s)<u_1<\cdots<u_k<\sigma(t)}
 \text{d}X^{i_1}_{\sigma^{-1}(u_1)}\cdots\text{d} X^{i_k}_{\sigma^{-1}(u_k)}
\]
\item\label{item:lin} \textbf{Linearity}: for all $I,J\in\mathcal{I}$ there exist a set
  $\mathcal{I}_{I,J}\subset\mathcal{I}$, such that
 \[
X^{I}_{s,t}X^{J}_{s,t} = \sum_{K\in\mathcal{I}_{I,J}}X^K_{s,t},
\]
that is any polynomial of iterated integrals can be represented by a
linear combination of iterated integrals (ref. \cite{SaintFlour}),

\item\label{item:sig} \textbf{Signature of linear paths}: if $X_t = a + bt$  for some
  $a,b\in\mathbb{R}^d$ and for all $t\in[0,T]$, moreover $I=(i_1,\dots,i_k)$, then
\[
X^{I}_{s,t} = \frac{(t-s)^k}{k!}\prod_{j=1}^kb_{i_k}.
\]

\item\label{item:mult} \textbf{Multiplicative property}: there exists a binary operation $\otimes$, such that for any
  $0\le s<t<u\le T$
\[
S_{s,t}(X)\otimes S_{t,u}(X) = S_{s,u}(X).
\]
The operation is specified in \cite{roughPaths98,SaintFlour}.

\item\label{item:uniq} \textbf{Uniqueness}: The main result of
  \cite{uniqueSig} is that
  $S_{s,t}(X)$ determines the function $u\mapsto X_u-X_s$ for
  $u\in[s,t]$ up to tree-like equivalence. A sufficient condition for
  the uniqueness is the existence of  $i\in\{1,\dots,d\}$ such that
  $X^i_u$ is strictly monotone increasing. 
 Moreover, typically, the first few terms of the signature contain
  most of the information on the path. 
\end{enumerate}
In particular, the uniqueness property (\ref{item:uniq}) highlights the relation between paths and
their signatures; the truncated signature of a path is regarded as a
projection of the path to a lower dimensional space.   
Furthermore, (\ref{item:lin}) enables us to apply linear regression based
machine learning techniques when estimating functionals of paths. 
Points (\ref{item:sig}) and (\ref{item:mult}) are relevant for the
computation of the truncated signatures of data streams. Finally,
(\ref{item:timeRep}) is used in section \ref{sec:quadraticVariation}, where we claim that
the lead-transform and the lag-transform both preserve the
signature of data streams. 

\subsection{Interpretations of the terms in the signature}\label{sec:interpretations}
The lower order terms in the signature have simple
interpretations. Recursively, one can derive the interpretation of
the higher order terms. Here, we focus on some terms of order at most
three. 

\paragraph{Increments.} For $i=1,\dots,d$ 
\[
X^{(i)}_{s,t}=X^i_t-X^i_s,
\]
that is the first order terms determine the increments of the
components. 

\begin{figure}[ht] 
\centering
\includegraphics[trim = 10mm 110mm 10mm 10mm,clip, width = 0.9\textwidth]{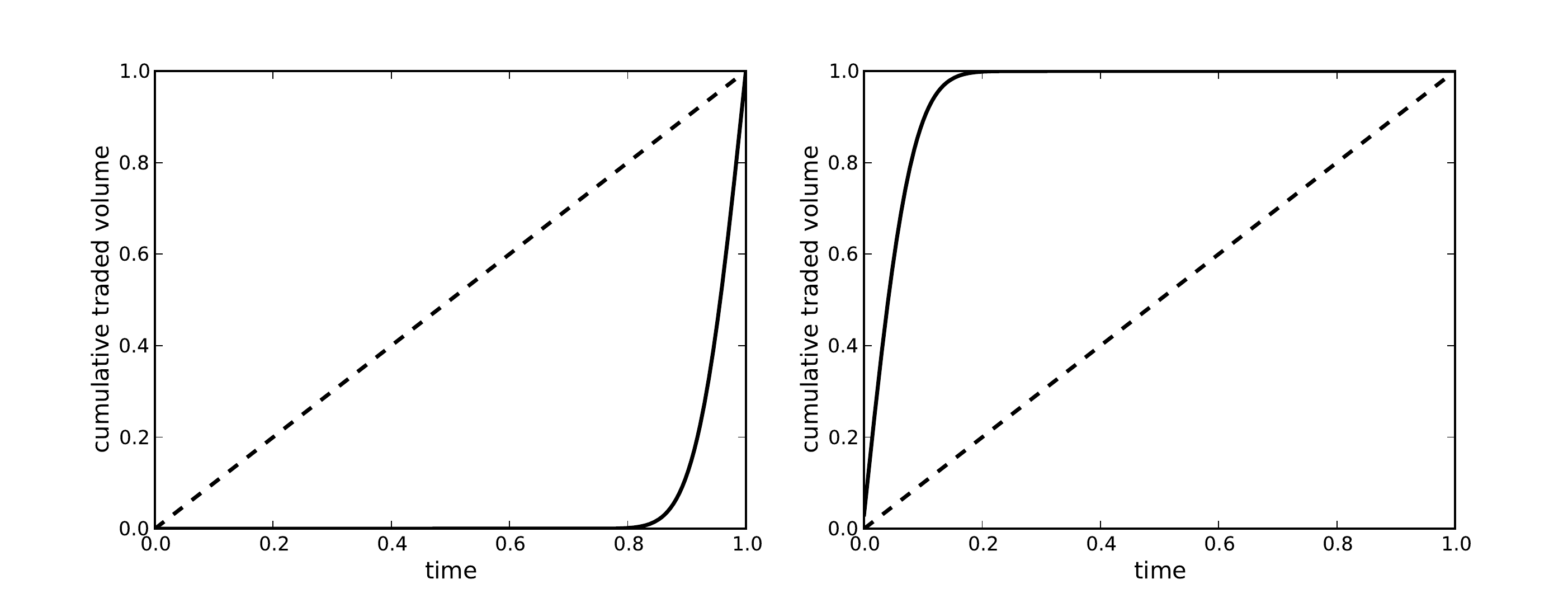}
\caption{Back-loaded and front-loaded volume profiles. }
\label{fig:path1}
\end{figure}
Figure \ref{fig:path1} shows two trajectories in two dimensions. The left-hand-side
plot, represents a volume profile in which the traded volume is
concentrated towards the end of the time interval, whereas on the
right-hand-side plot the traded volume is concentrated at the
beginning of the time interval. Both curves have unit increments in
both of their time and cumulative volume components. 

\paragraph{Higher order terms.} Higher order iterated integrals can be
interpreted as generalised polynomials of paths. Linear regression on
these generalised polynomials can be interpreted as polynomial
regression directly on the input paths. In order to obtain better
accuracy in machine learning, we explore the relevance of each of
these terms. 

However,  not all terms are required for the unique representation of a path in terms of its iterated
integrals. In particular, some of the
iterated integrals are determined by other iterated integrals as the
following examples show.
\begin{align*}
X^{(i,i)}_{s,t} & = \tfrac{(X^{(i)}_{s,t})^2}{2} \text{ for }
i\in\{1,\dots,d\} \\
X^{(i,j)}_{s,t} & = X^{(i)}_{s,t}X^{(j)}_{s,t} - X^{(j,i)}_{s,t}
\text{ for } i,j\in\{1,\dots,d\}
\end{align*}
These identities are implied by property (\ref{item:lin}) in section
\ref{sec:intuitiveIntroduction}. The minimal set of information that
uniquely determines all the iterated integrals of a path through
non-linear operations is called the \emph{log-signature} of the path
(ref. \cite{roughPaths98,SaintFlour,FrizVictoir}). The components of
the log-signature consist of  certain linear combinations of the
iterated integrals. We briefly introduce two of these terms -- the
\emph{area} and a \emph{second order area} -- below.

\begin{figure}[ht] 
\centering
\includegraphics[trim = 10mm 110mm 10mm 10mm,clip, width = 0.9\textwidth]{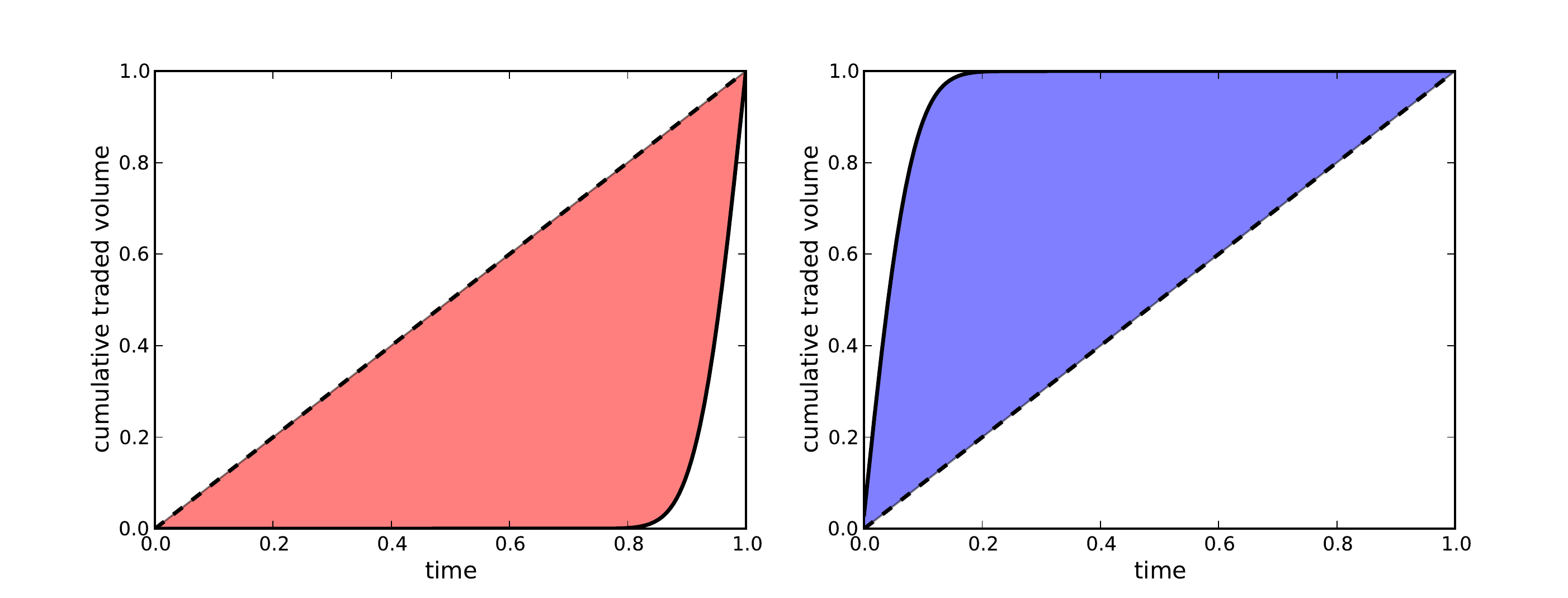}
\caption{Back-loaded and front-loaded volume  profiles with shaded
  signed area}
\label{fig:path1a}
\end{figure}

\paragraph{Area.} Now, we give a possible interpretation of certain linear combinations
of the higher order
iterated integrals. 
For $1\le i< j\le d$, the terms
\[
A^{i,j}_{s,t}:=\frac{1}{2}\left(
\int_{s<u_1<u_2<t}\text{d}X^1_{u_1}\text{d}X^2_{u_2}-
\int_{s<u_1<u_2<t}\text{d}X^2_{u_1}\text{d}X^1_{u_2}
\right)=\frac{1}{2}(X^{(i,j)}_{s,t}-X^{(j,i)}_{s,t})
\]
determine the signed area between the curve
$u\mapsto(X^i_u,X^j_u)$ for $u\in[s,t]$ and the cord that connects the
points $(X^i_s,X^j_s)$ and $(X^i_t,X^j_t)$. For example the blue
(respectively red) shaded
area in figure \ref{fig:path1a} indicates negative (positive) area. As
noted above, the plotted curves have unit increments in both of their
components, however one has a relatively large positive area, whereas
the other has a relatively large negative area. 

\begin{figure}[ht] 
\centering
\subfigure[Volume profiles with zero total area over \unitInt]{
\includegraphics[trim = 10mm 110mm 10mm 10mm, clip,width =
0.9\textwidth]{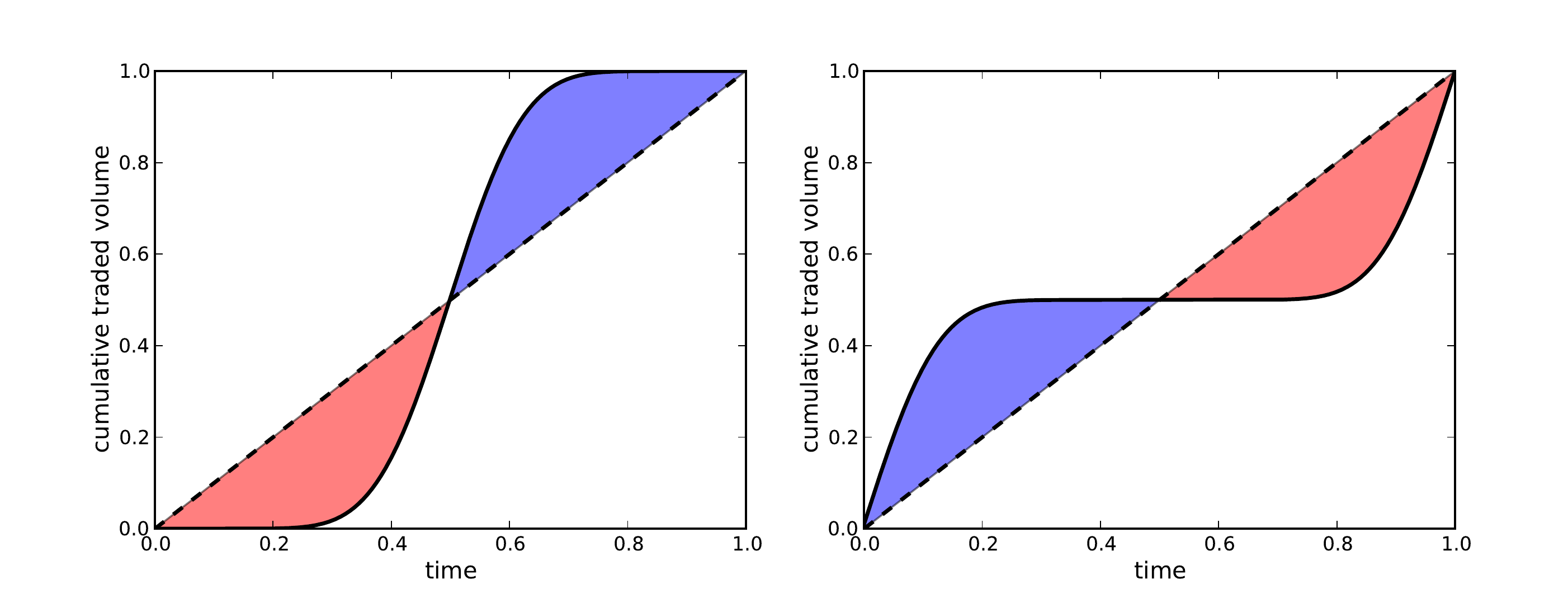}}
\subfigure[The areas as function of time]{
\includegraphics[trim = 10mm 10mm 10mm 120mm, clip,width =
0.9\textwidth]{plot_area2}
}
\caption{Mid-loaded (on the left) and front-and-back-loaded (on the
  right) volume profiles}
\label{fig:path2}
\end{figure}

\paragraph{Second order area.} The upper graphs of figure
\ref{fig:path2} show two paths in two dimensions -- the
mid-loaded and the front-and-back-loaded volume profiles. Both trajectories have unit
increments in both their components, and both have zero signed
areas. In order to tell the difference between these two curves
through their signature, one has to look into the third order
terms. In particular, on the bottom graphs in figure \ref{fig:path2} 
the curve  $u\mapsto (X^1_u,A^{1,2}_{0,u})$ is plotted; the area under this curve
is negative for the mid-loaded volume profile, and positive for the
front-and-back-loaded volume profile. 
This second order area is given by the formula
\[
A^{1,(1,2)}_{s,t}:= \frac{1}{2}\left(
\int_{s<u_1<u_2<t}\text{d}X^1_{u_1}\text{d}A^{1,2}_{s,u_2}-
\int_{s<u_1<u_2<t}\text{d}A^{1,2}_{s,u_1}\text{d}X^1_{u_2}
\right)=\frac{1}{2}(X^{(1,1,2)}_{s,t}-X^{(1,2,1)}_{s,t}),
\]
where the last step is implied by property  (\ref{item:lin})
(the reader is referred to section 2.2.3 of \cite{SaintFlour} for
a rigorous proof). 

Using similar recursive arguments, one can identify higher order areas
and represent them in terms of linear combinations of iterated integrals. 

\paragraph{Lead-lag relationship.} As the examples of this section show, the
terms in the signature capture the lead-lag relationship. In
particular, if an increase (respectively decrease) of the component $X^1$ is typically
followed by an increase (decrease) in the component $X^2$, then the area
$A^{1,2}$ is positive. If a move in $X^1$ is followed by a move in
$X^2$ to the opposite direction, the area is negative.

\subsection{Signature of data streams}
The previous section introduced the signature of continuous
trajectories. In financial time series analysis, one can observe
data streams, i.e. trajectories defined at and indexed by finitely many time points:
$(\widehat{X}_{t_i})_{i=0}^N$ in $\mathbb{R}^d$. 

A possible way to define the signature of a data stream is via the
iterated integrals of its the piece-wise linear interpolation. That
is, we introduce the continuous path
\[
X_u = \widehat{X}_{t_i} +
\frac{u-t_i}{t_{i+1}-t_i}(\widehat{X}_{t_{i+1}}-\widehat{X}_{t_i}) \text{
  for } u \in [t_i,t_{i+1}],
\]
%

%
 and we define \emph{signature of the stream} $(\widehat{X}_{t_i})_{i=0}^N$  as
 \[
 S_{t_0,t_N}(\widehat{X}):=S_{0,2N}(X). 
 \]

We note that Ni and Lyons \cite{NiLyons} define the signature of data
streams by constructing axis paths instead of using piece-wise linear interpolation. 

\subsection{Quadratic variation}\label{sec:quadraticVariation}
We have shown that the terms of the signature -- the iterated
integrals -- measure various path-dependent quantities. However, the
quadratic variation of the process is not directly captured. Since
this quantity has a high relevance in financial applications, we have
found it crucial to incorporate quadratic variation into the
signature. This is possible through the \emph{lead-lag
  transformation} of streams. 

In particular, given a stream  $(\widehat{X}_{t_i})_{i=0}^N$ in
$\mathbb{R}^d$, we define its \emph{lead-transformed} stream $(\widehat{X}^{\text{lead}}_j)_{j=0}^{2N}$ by
\[
\widehat{X}^{\text{lead}}_j=\left\{
\begin{array}{ll}
\widehat{X}_{t_i} & \text{ if } j = 2i \\
\widehat{X}_{t_{i}} & \text{ if } j = 2i-1 
\end{array}
\right.
\]
Moreover, we define its \emph{lag-transformed} stream $(\widehat{X}^{\text{lag}}_j)_{j=0}^{2N}$ by
\[
\widehat{X}^{\text{lag}}_j=\left\{
\begin{array}{ll}
\widehat{X}_{t_i} & \text{ if } j = 2i \\
\widehat{X}_{t_{i}} & \text{ if } j = 2i+1 
\end{array}
\right.
\]
Finally, the \emph{lead-lag-transformed} stream takes values in
$\mathbb{R}^{2d}$ and is defined by the paired stream
\[
(\widehat{X}^{\text{lead-lag}}_j)_{j=0}^{2N} = (\widehat{X}^{\text{lead}}_j,\widehat{X}^{\text{lag}}_j)_{j=0}^{2N}.
\]

Note that the axis path $X^{\text{lead}}$ corresponding to the
lead-transform stream is a time-reparameterisation of the axis path
$X$, hence 
\[
S_{t_0,t_N}(\widehat{X}) = S_{0,2N}(\widehat{X}^{\text{lead}}),
\]
and similarly 
\[
S_{t_0,t_N}(\widehat{X}) = S_{0,2N}(\widehat{X}^{\text{lag}}).
\]
Furthermore, the (signed) area between the $i$th component of the
lead-transform and the $j$th component of the lag-transform equals to
the quadratic cross-variation of the trajectories $\widehat{X}^i$ and
$\widehat{X}^j$.

In practice, we use a partial lead-lag-transform, that is we take the
lead-transform of the input stream, and pair it with the lag-transform
of only those components of which the quadratic variation is
relevant (see section \ref{sec:inputData}). 

\begin{figure}[ht] 
\centering
\subfigure[Lead-lag transformed path as function of time]{
\includegraphics[trim = 10mm 5mm 194mm 15mm, clip,width =
0.47\textwidth]{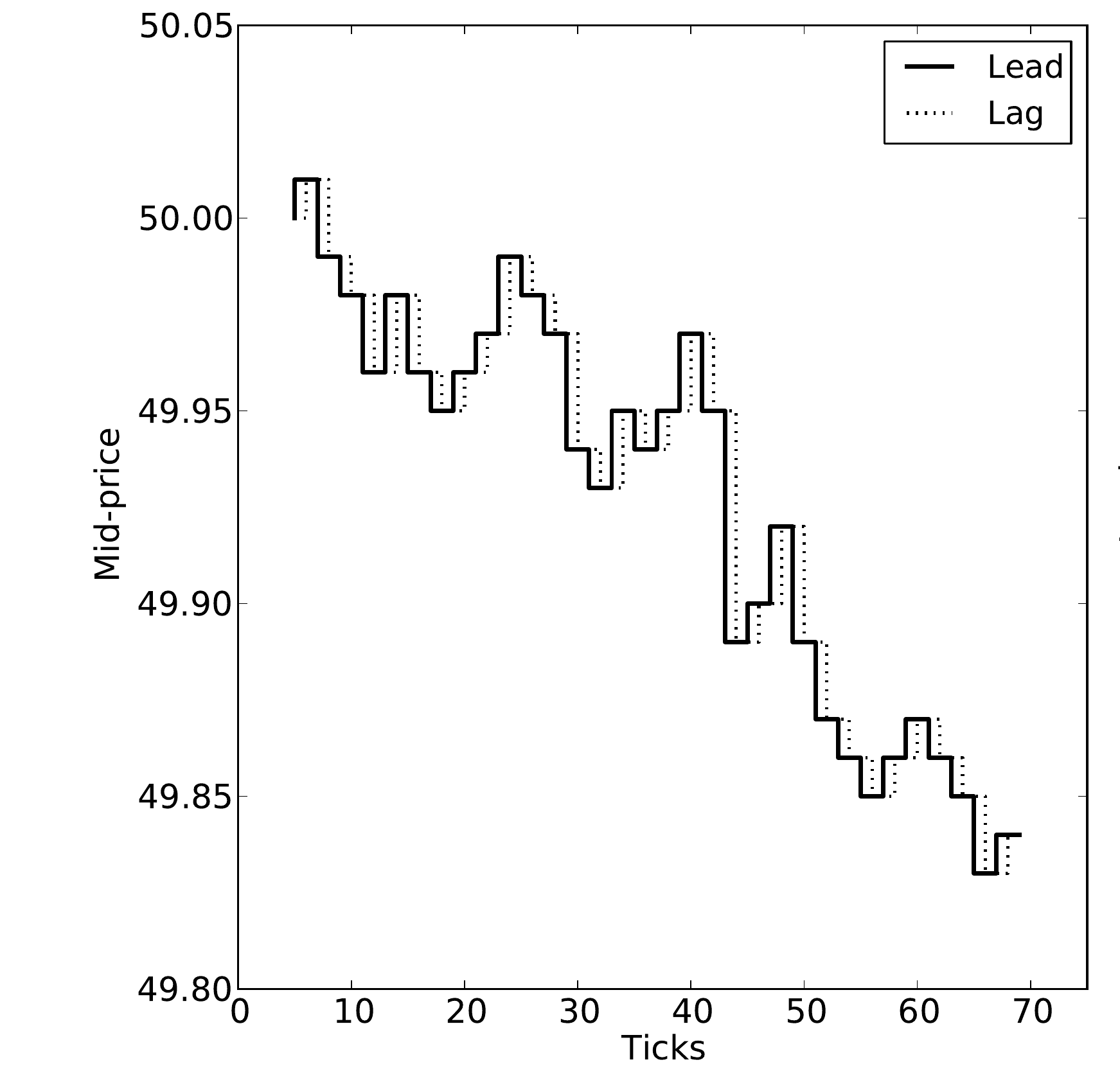}
}
\subfigure[Lead-lag transformed path in the lead-lag space]{
\includegraphics[trim = 185mm 5mm 19mm 15mm, clip,width =
0.47\textwidth]{plot_leadLag19}
}
\caption{Lead-lag transformed mid-price}
\label{fig:leadLag}
\end{figure}

For instance, in figure \ref{fig:leadLag}, two representations of the axis path
corresponding to the lead-lag
transform of a simulated mid-price stream are shown. The signed area
between the curve and the cord on the right-hand side plot
\emph{equals to the quadratic variation} of the mid price.

\section{Methodology}\label{sec:methodology}

\subsection{Input data}\label{sec:inputData}
Our numerical tests are focused on two futures markets: WTI crude oil (NYMEX)
and FTSE 100 index (NYSE Liffe). We work with sub-sampled level-one
order book data of the front-month
(closest maturity) futures as raw data, in particular:
\begin{itemize}
\item $P^a$: best ask price, 
\item $P^b$: best bid price,
\item $V^a$: number of outstanding orders at best ask price,
\item $V^b$ number of outstanding orders at best bid price,
\item $C$: cumulative traded volume. 
\end{itemize}
The sampling frequency is specified in the description of the
particular tests. 

In each test, the input variables are streams of
level-one order book data; the features used by the classification
algorithm are the terms of the truncated signatures of these streams. The lengths of the input streams depend on
the particular test. However, in each case, we transform and normalise the
data as follows. For each stream
$\widehat{Y}:=((P^a,P^b,V^a,V^b,C^t)_{t_i})_{i=0}^N$, we define the components:
\begin{itemize}
\item the \emph{normalised time}: $u_{t_i} := \tfrac{t_i-t_0}{t_N-t_0}$, 

\item the \emph{normalised logarithm of mid-price}: 
\[
p_{t_i} := C_p\log\frac{P^a_{t_i}+P^b_{t_i}}{2},
\]
where for each stream, $C_p$ is chosen to satisfy $\text{StDev}(\Delta
p)=1$, 

\item the \emph{standardised spread}: $s_{t_i} :=
  C_s(P^a_{t_i}-P^b_{t_i})$ where for each stream, $C_s$ is chosen to satisfy
  $\text{StDev}(s)=1$,

\item the \emph{normalised imbalance} $d_{t_i} :=
  \tfrac{V^a_{t_i}-V^b_{t_i}}{V^a_{t_i}+V^b_{t_i}}$,

\item the \emph{normalised cumulative volume}: $c_{t_i} :=
  \tfrac{C_{t_i}}{C_{t_N}}$.
\end{itemize}
Using these components and the time stamps, each $\widehat{Y}$ stream is transformed to
$\widehat{X}:=(u_{t_i},p_{t_i},s_{t_i},d_{t_i},c_{t_i})_{i=0}^N$. 

In order to capture the quadratic variation of the log-price, the
lead-transform $\widehat{X}^{\text{lead}}$ is computed and is paired
  with the lag-transform of $p$; hence, the input streams are of the form:
\begin{equation}
(\widehat{Z}_{s_i})_{i=0}^{2N} := \Big((u^{\text{lead}}_{s_i},p^{\text{lead}}_{s_i},s^{\text{lead}}_{s_i},d^{\text{lead}}_{s_i},c^{\text{lead}}_{s_i},
p^{\text{lag}}_{s_i})\Big)_{i=0}^{2N}.
\label{eq:inputData}
\end{equation}
We compute the signature of each stream up to depth $4$, that is for
all iterated integrals that are indexed with multi-indices of length
at most $4$. The signatures are computed with the \verb_sigtools_ Python package which is based
on the the \verb_libalgebra_ library of the CoRoPa
project\footnote{Version 0.3, ref.: \url{http://coropa.sourceforge.net/}}.

\subsection{Linear regression-based classification}
In each numerical test (ref. sections \ref{sec:timeBucket},
\ref{sec:AvsB} and \ref{sec:localAvsB}) a set of input streams is given, which we
 partition into two categories along some properties of the streams.
 The aim of the tests is to
predict the category of a stream using its signature, that is when the independent
variables -- or features --  are the terms in the truncated signatures of the input
streams. A certain
percentage of the streams in each category is used as a learning set and the remaining
streams are used for out-of-sample testing. 
The value of the dependent variable -- or response -- is defined to be $0$ along the
streams of one category and $1$ along the streams of the other
category. The classification method is  linear regression
combined with LASSO (least absolute shrinkage and selection operator)
shrinkage (ref. \cite{lasso}) in order to identify the relevant terms
in the signature. That is, we minimise the following objective
function:
\[
\min_\beta \left[\sum_{k=1}^K \left( \sum_{I\in\mathcal{I}_m} \beta_I
  Z^{I}_{0,2N_k}(k) - y(k) \right)^2 + \alpha \sum_{I\in\mathcal{I}_m} |\beta_I|\right],
\]
where $\mathcal{I}_m$ denotes the set of multi-indices of length at most
$m$, $\beta = (\beta_I)_{I\in\mathcal{I}_m}$, $K$ denotes the
cardinality of the
learning set, $Z(k)$ denotes the axis
path corresponding to the $k$th input stream $(\widehat{Z}_i)_{i=0}^{2N_k}(k)$, $y(k)$ denotes the
category ($0$ or $1$) of the $k$th input stream, and $\alpha$ is the
shrinkage parameter. For each test case, the level of $\alpha$ is
determined by $k$-fold cross-validation.

Since the LASSO shrinkage is sensitive to the rescaling of variables, we
centralise and normalise each independent variable across streams
based on first and second moments that are estimated using the
learning set of streams. 

 We used the
\verb_scikit-learn_ Python
package\footnote{Version 0.14, ref.: \url{http://scikit-learn.org/}} to compute the
LASSO-based regression coefficients.  

\subsection{Statistical indicators}\label{sec:statInd}
In order to measure the accuracy of the classification, for each
 test case we compute the following measures based on the
regressed values. 
\begin{enumerate}[(i)]
\item\label{item:KSInSample} The Kolmogorov-Smirnov distance of the distributions of regressed
  values for each category computed using the learning set of streams. 

\item The Kolmogorov-Smirnov distance of the distributions of regressed
  values for each category computed using the out-of-sample  streams. 

\item Setting the decision boundary at the level where the
  Kolmogorov-Smirnov distance is attained\footnote{This choice of
    threshold maximises the ratio of correct predictions.} in (\ref{item:KSInSample}),
  the number of true positives, false positives, true negatives, false
  negatives and the ratio of correct classification are computed. 

\item\label{item:ROCInSample} We plot the receiver operating characteristic (ROC) curve and
  compute the area under the ROC curve for the regressed values
  calculated using the learning set of streams.  

\item We repeat (\ref{item:ROCInSample}) using the score values computed
  using the out-of-sample streams. 
\end{enumerate}

\section{Numerical tests}\label{sec:numericalResults}
In this section, we describe various test cases and the numerical
results.

\subsection{Characterising standard time buckets of a futures market}\label{sec:timeBucket}
As an initial exercise, we explore to what extent the signatures of
30-minute data streams determine the time bucket they are sampled
from. In particular, we run binary classification on various pairs of
time intervals, and explore which time buckets show similar behaviour
and which time buckets have unique characteristics. 

Since there are recurring patterns in the market activity, the data
 standardisation described in section \ref{sec:inputData} is done in
 order to \emph{hide the obvious} features. Primarily, the order of magnitude of
 the trading volume or the level of volatility are all 
discarded
 through the standardisation process. The information that remains is
 what is captured by the first and higher order areas (see section
 \ref{sec:interpretations}). 

\paragraph{Input streams} The underlying instrument is the front-month WTI
crude oil future (NYMEX). The input streams are sampled by the minute from standard 30-minute
time intervals on each trading date between $1$st January 2011 and $1$st
February 2013. From each time-bucket $75\%$ of the streams are
randomly selected for learning and the remaining $25\%$ are used for
out of sample testing. 

\paragraph{14:00-14:30} Despite hiding the obvious features, the
14:00-14:30 EST time bucket shows very distinctive characteristics. Since
the open outcry trading closes at 14:30 EST, this result -- after all -- is not
surprising. 

\begin{figure}[ht] 
\centering
\subfigure[\textbf{Learning
set}: Estimated densities of the regressed values, K-S distance: $0.9$,
correct classification: $95\%$]{
\includegraphics[trim = 10mm 5mm 154mm 15mm, clip,width =
0.47\textwidth]{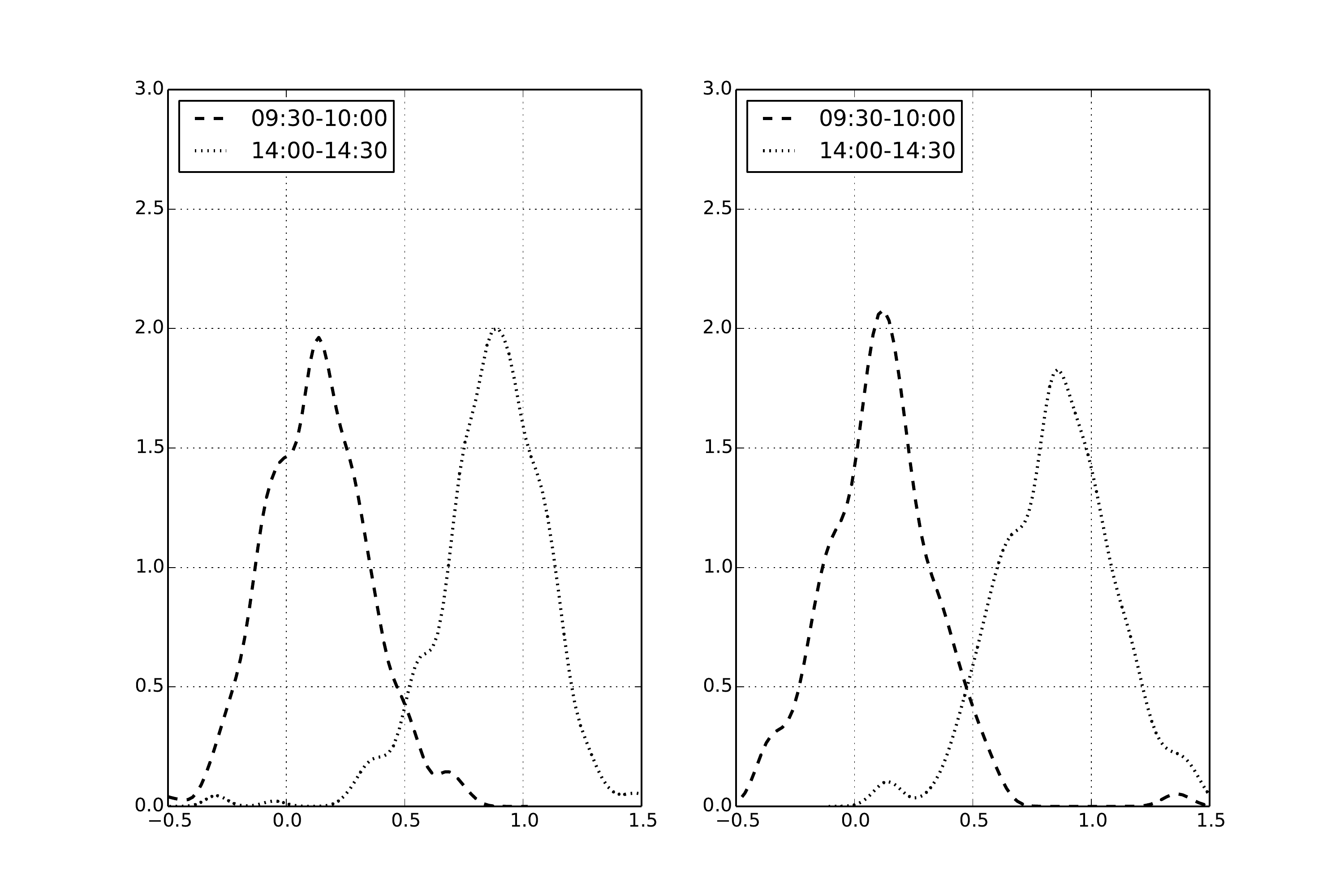}
}
\subfigure[\textbf{Out of
sample}: Estimated densities of the regressed values, K-S distance:
$0.91$, correct classification: $95\%$]{
\includegraphics[trim = 154mm 5mm 10mm 15mm, clip,width =
0.47\textwidth]{CLN_NOV_plot_CDF_2011_2012_2013_930vs1400_PDF}
}
\subfigure[\textbf{ROC curve.} Area under ROC -- learning set: 0.986, out of
sample: 0.984 ]{
\includegraphics[trim = 10mm 3mm 10mm 10mm, clip,width =
0.8\textwidth]{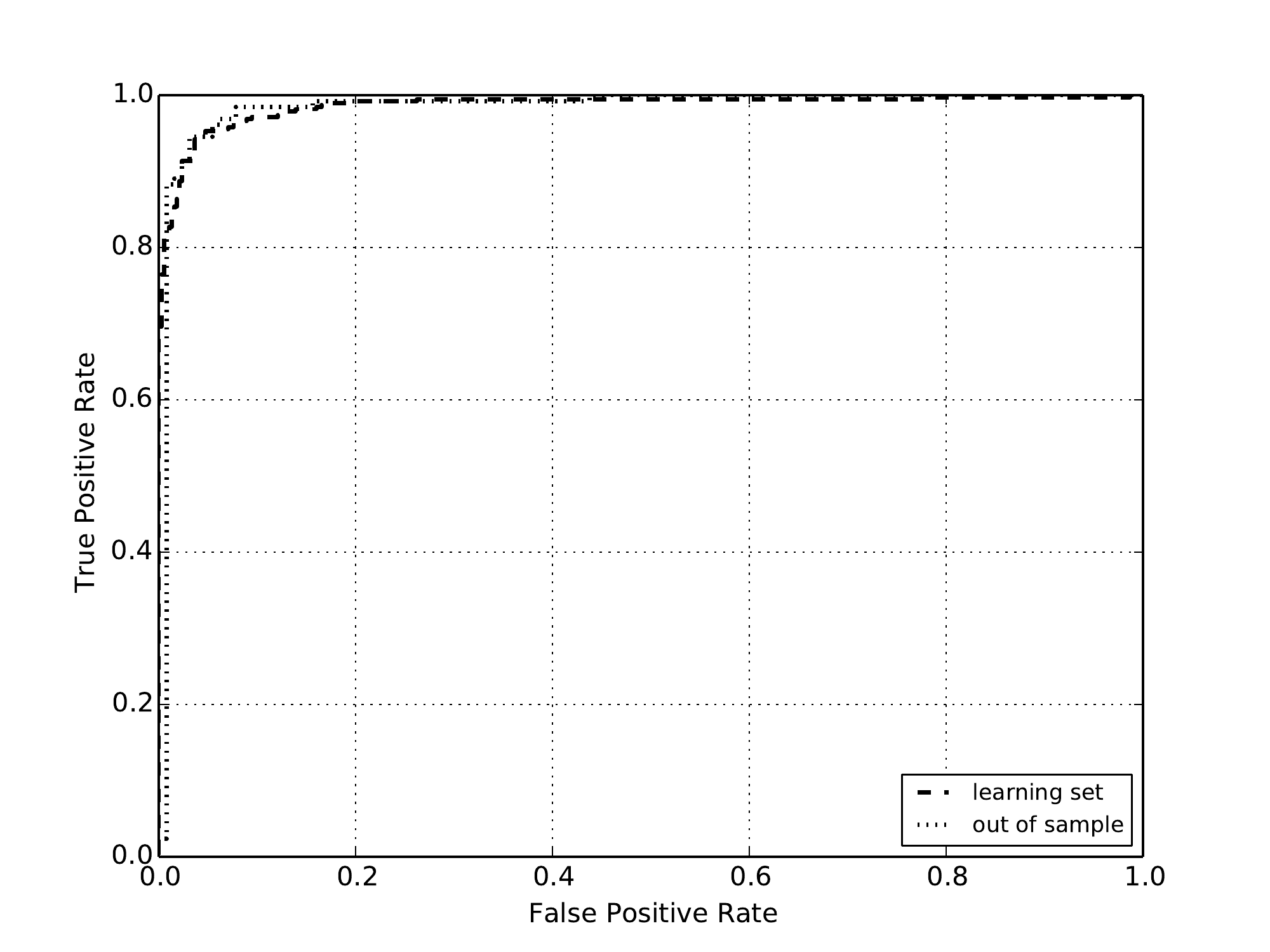}
}
\caption{14:00-14:30 EST versus 9:30-10:00 EST}
\label{fig:140vs930}
\end{figure}

\begin{figure}[ht] 
\centering
\subfigure[\textbf{Learning
set}: Estimated densities of the regressed values, K-S distance: $0.8$, correct classification: $90\%$]{
\includegraphics[trim = 10mm 5mm 154mm 15mm, clip,width =
0.47\textwidth]{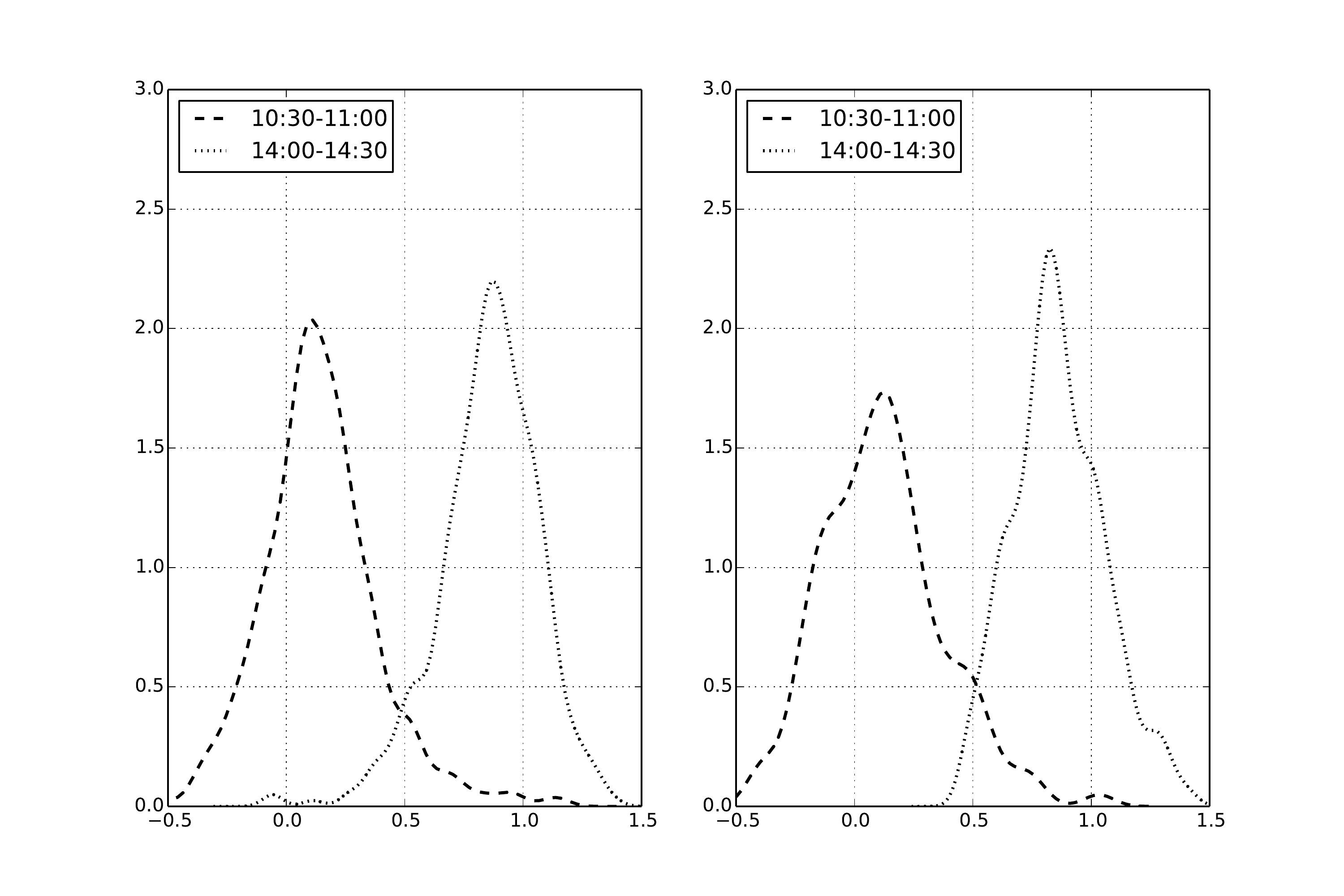}
}
\subfigure[\textbf{Out of
sample}: Estimated densities of the regressed values, K-S distance:
$0.84$, correct classification: $89\%$]{
\includegraphics[trim = 154mm 5mm 10mm 15mm, clip,width =
0.47\textwidth]{CLN_NOV_plot_CDF_2011_2012_2013_1030vs1400_PDF}
}
\subfigure[\textbf{ROC curve.} Area under ROC -- learning set: 0.976, out of
sample: 0.986 ]{
\includegraphics[trim = 10mm 3mm 10mm 10mm, clip,width =
0.8\textwidth]{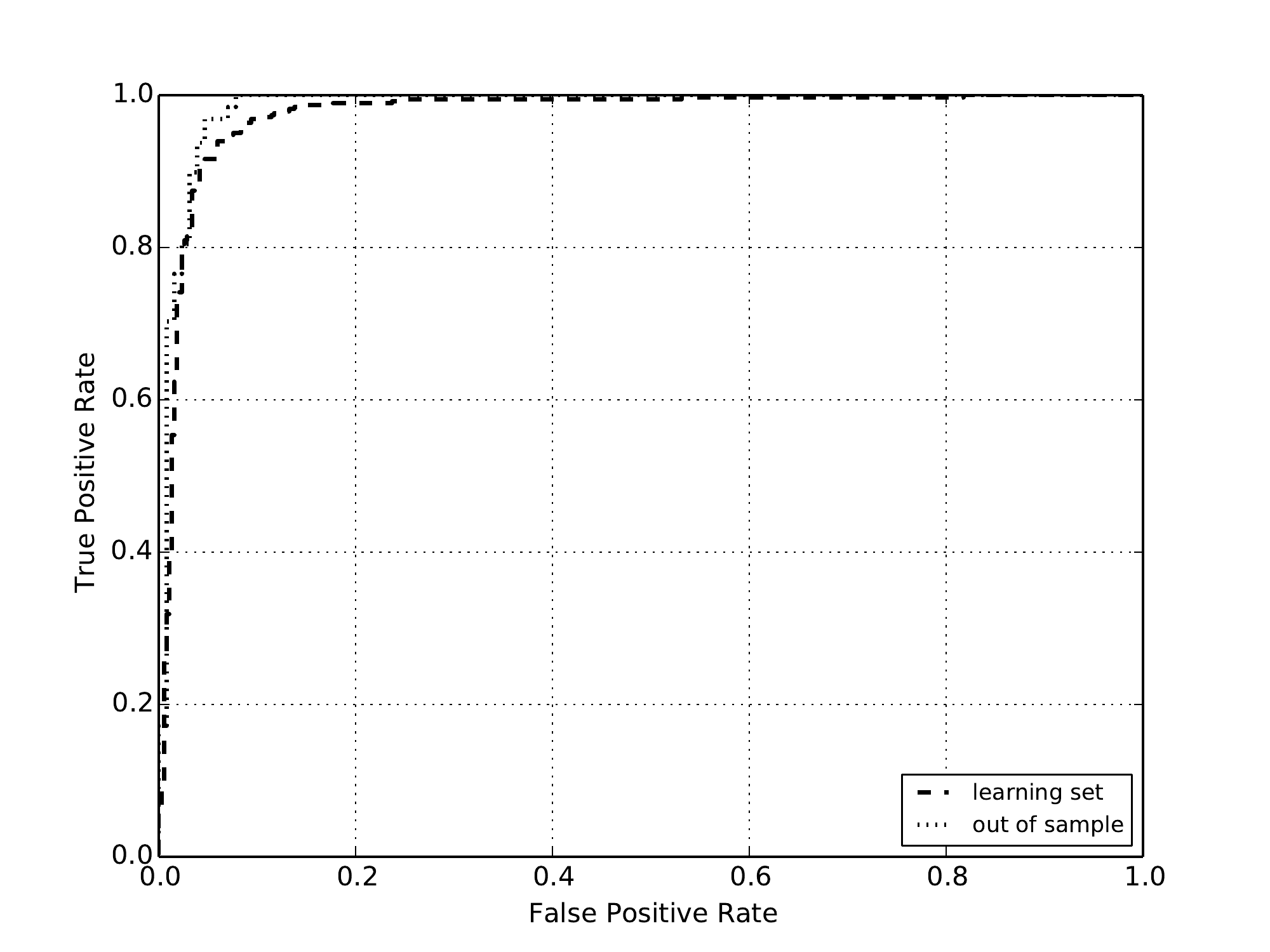}
}
\caption{14:00-14:30 EST versus 10:30-11:00 EST}
\label{fig:140vs1030}
\end{figure}

In figure \ref{fig:140vs930}, we present the classification results
(the estimated densities of regressed values and the ROC curve) when we
compare the 9:30-10:00 time bucket to 14:00-14:30 time bucket. 
Similarly, in figure \ref{fig:140vs1030}, the results of the binary classification of 
the 10:30-11:00 and 14:00-14:30 intervals are shown. 

\begin{table}
\centering
\begin{tabular}{|c|cc|cc|cc|}
\hline
                & \multicolumn{2}{c|}{Kolmogorov-Smirnov}
               & \multicolumn{2}{c|}{Area under}  & \multicolumn{2}{c|}{Ratio
               of correct } \\
                &  \multicolumn{2}{c|}{ distance}
               &  \multicolumn{2}{c|}{ROC curve}  &
               \multicolumn{2}{c|}{  classification} \\
 bucket  & LS & OS & LS &
OS &  LS & OS \\
\hline
9:30-10:00 & 0.906 & 0.914 & 0.986 & 0.984 & 0.952 & 0.946 \\

10:30-11:00 & 0.880 & 0.922 & 0.976 & 0.986 & 0.939 & 0.946 \\

11:30-12:00 & 0.940 & 0.961 & 0.995 & 0.992 & 0.969 & 0.973 \\

12:30-13:00 & 0.888 & 0.906 & 0.986 & 0.988 & 0.942 & 0.950 \\

13:30-14:00 & 0.882 & 0.914 & 0.979 & 0.974 & 0.940 & 0.942\\

\hline
\end{tabular}
\caption{The 14:00-14:30 EST time bucket compared some other intervals. \\
  ``LS'': learning set, ``OS'': out-of-sample set.}\label{tab:140}
\end{table}

\begin{figure}[ht] 
\centering
\includegraphics[trim = 7mm 3mm 10mm 10mm, clip,width =
0.8\textwidth]{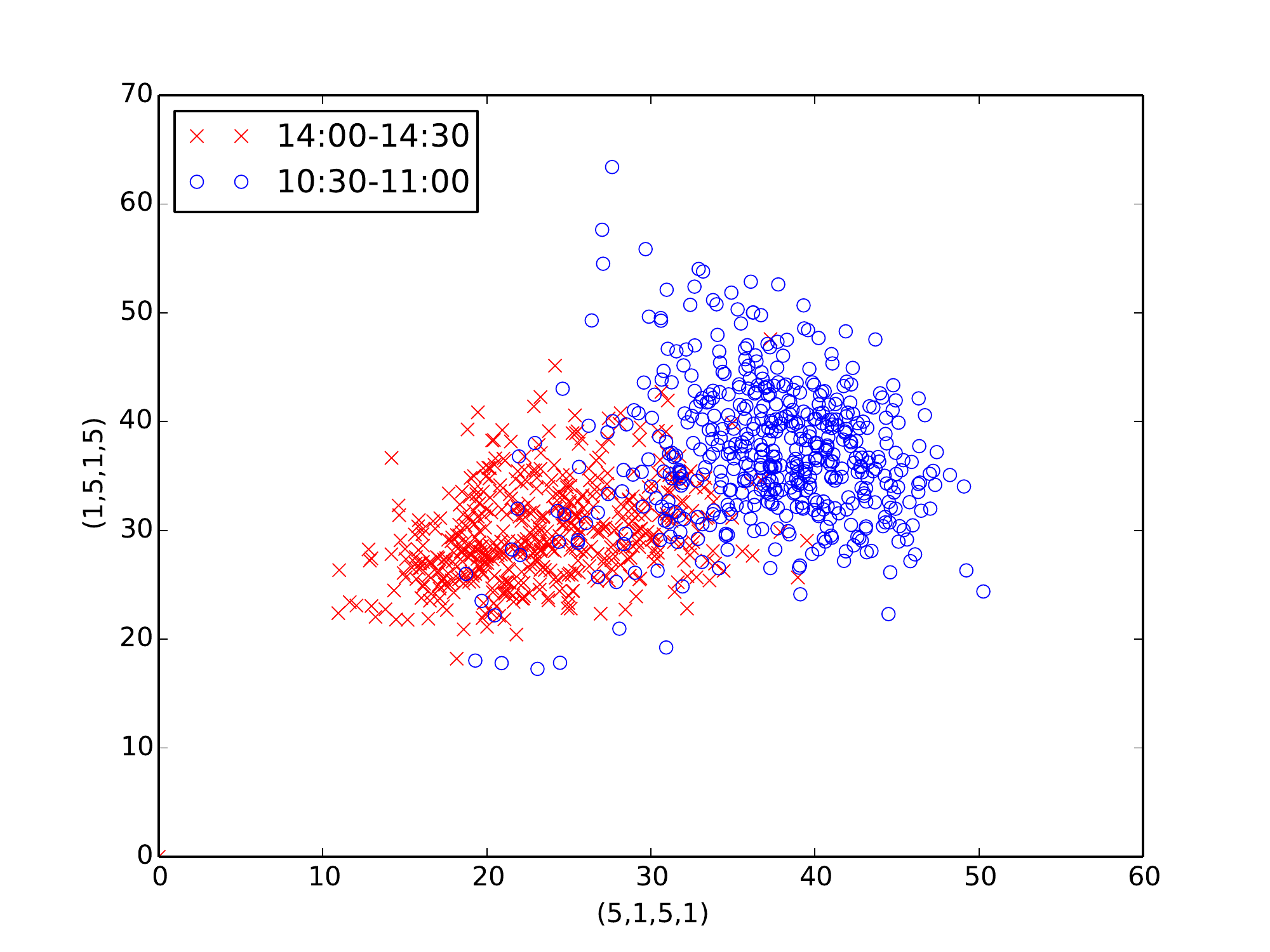}
\caption{Projection of the signatures (i) - 14:00-14:30 versus 10:30-11:00}
\label{fig:140vs1030Proj1}
\end{figure}

\begin{figure}[ht] 
\centering
\includegraphics[trim = 7mm 3mm 10mm 10mm, clip,width =
0.8\textwidth]{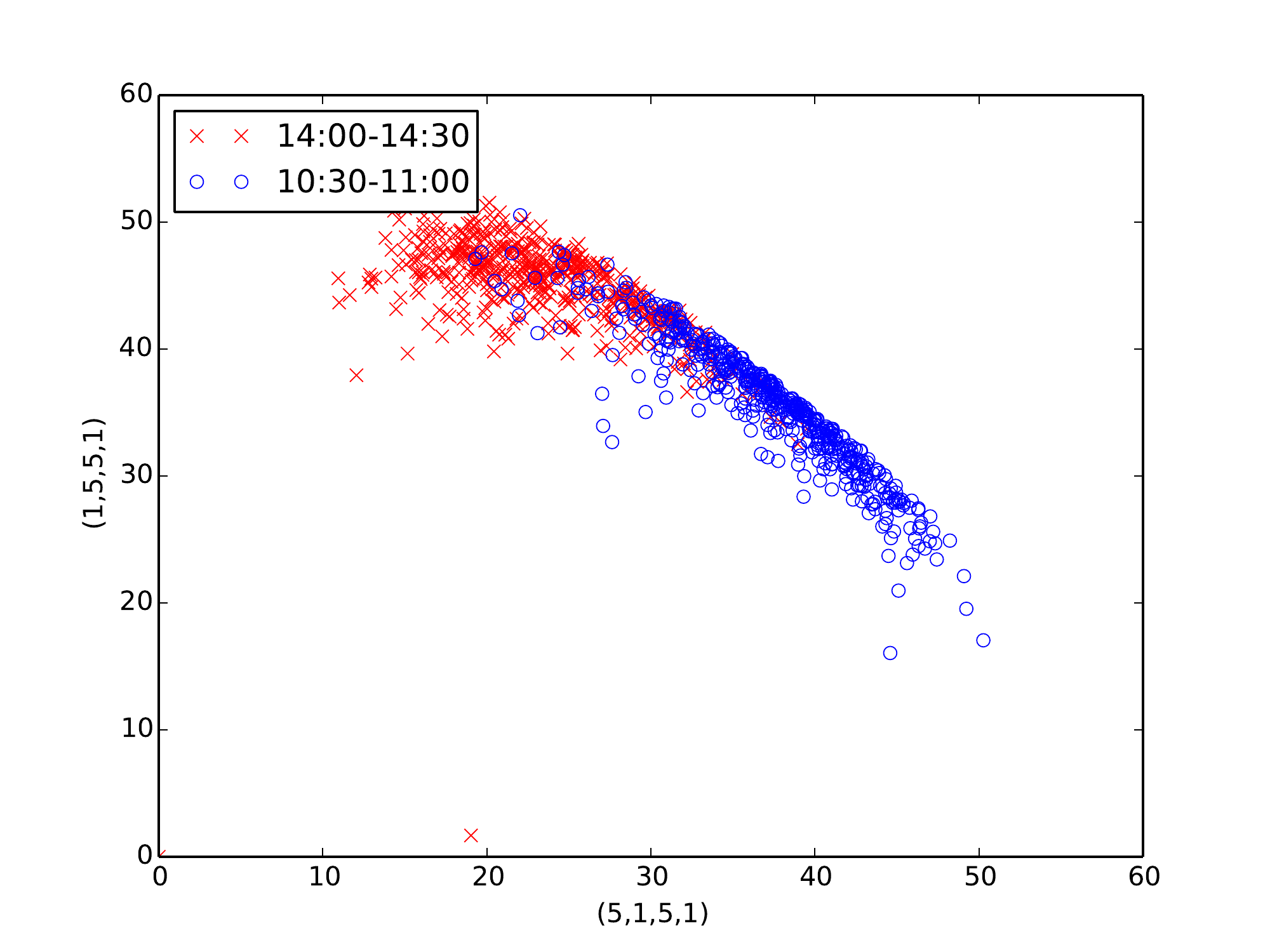}
\caption{Projection of the signatures (ii) - 14:00-14:30 versus 10:30-11:00}
\label{fig:140vs1030Proj2}
\end{figure}

\begin{figure}[ht] 
\centering
\includegraphics[trim = 7mm 3mm 10mm 10mm, clip,width =
0.8\textwidth]{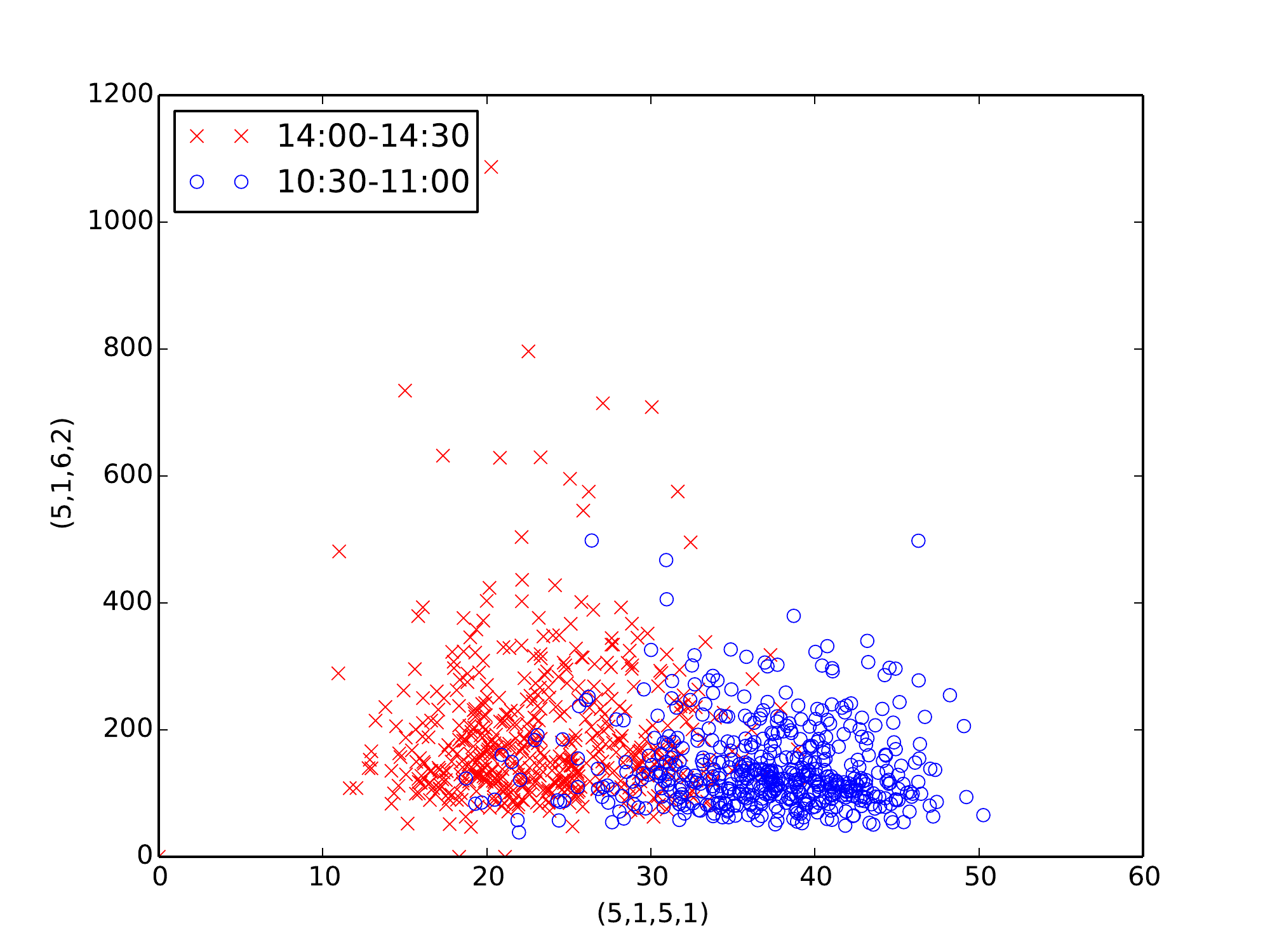}
\caption{Projection of the signatures (iii) - 14:00-14:30 versus 10:30-11:00}
\label{fig:140vs1030Proj3}
\end{figure}

Table \ref{tab:140} contains some more results. Apart from the
Kolmogorov-Smirnov distance, area under ROC curve and ratio of correct
classification, the
table shows the number of relevant variables identified by the LASSO
shrinkage algorithm. In particular, the most relevant terms are fourth
order iterated integrals such as the ones that correspond to these multi-indices:
\[
(1,5,1,5), \ (5,1,5,1), \ (1,5,5,1), \text{ and } (5,1,6,2),
\]
where $1$, $2$, $5$ and $6$ are the indices of $t^{\text{lead}}$,
$p^{\text{lead}}$, $c^{\text{lead}}$ and $p^{\text{lag}}$
respectively. The first three terms are involved in the higher order
characteristics of the cumulative volume profile. The last term includes the lead-lag-transformed log-mid
price, hence it is related to the quadratic variation of the
mid-price. In figures
\ref{fig:140vs1030Proj1}, \ref{fig:140vs1030Proj2} and \ref{fig:140vs1030Proj3}, the signatures of the input streams are
projected on to two-dimensional subspaces that are spanned by pairs of
the most relevant features. The projections demonstrate how
 the particular fourth order terms capture the difference between the
 two time buckets.

\paragraph{11:30-12:00} As another example, we also discuss how the
11:30-12:00 EST time bucket is distinguished from the other time-buckets.

\begin{table}
\centering
\begin{tabular}{|c|cc|cc|cc|}
\hline
                & \multicolumn{2}{c|}{Kolmogorov-Smirnov}
               & \multicolumn{2}{c|}{Area under}  & \multicolumn{2}{c|}{Ratio
               of correct } \\
                &  \multicolumn{2}{c|}{ distance}
               &  \multicolumn{2}{c|}{ROC curve}  &
               \multicolumn{2}{c|}{  classification} \\
 bucket  & LS & OS & LS &
OS &  LS & OS \\
\hline
9:30-10:00 & 0.621 & 0.609 & 0.884 & 0.846 & 0.810 & 0.798 \\

10:30-11:00 & 0.455 & 0.461 & 0.800 & 0.782 & 0.727 & 0.713 \\

10:30-11:00 & 0.455 & 0.461 & 0.800 & 0.782 & 0.727 & 0.713 \\

12:30-13:00 & 0.399 & 0.313 & 0.760 & 0.688 & 0.699 & 0.636\\

14:00-14:30 & 0.940 & 0.961 & 0.995 & 0.992 & 0.969 & 0.973 \\

\hline
\end{tabular}
\caption{The 11:30-12:00 EST time bucket compared some other intervals.\\
``LS'': learning set, ``OS'': out-of-sample set.}\label{tab:113}
\end{table}

The results are summarised in table \ref{tab:113}. The 11:30-12:00
time bucket shows more similarities with the reference time intervals,
however it is still distinguishable, although more variables are
required. Similarly as in the previous case, all the relevant
variables correspond to certain fourth order iterated integrals; apart from
the higher order characteristics of the cumulative volume profile and
the quadratic variation, iterated integrals that contain the spread are
also identified. 

\begin{figure}[ht] 
\centering
\subfigure[\textbf{Learning
set}: Estimated densities of the regressed values, K-S distance: $0.29$,
correct classification: $55\%$]{
\includegraphics[trim = 10mm 5mm 154mm 15mm, clip,width =
0.47\textwidth]{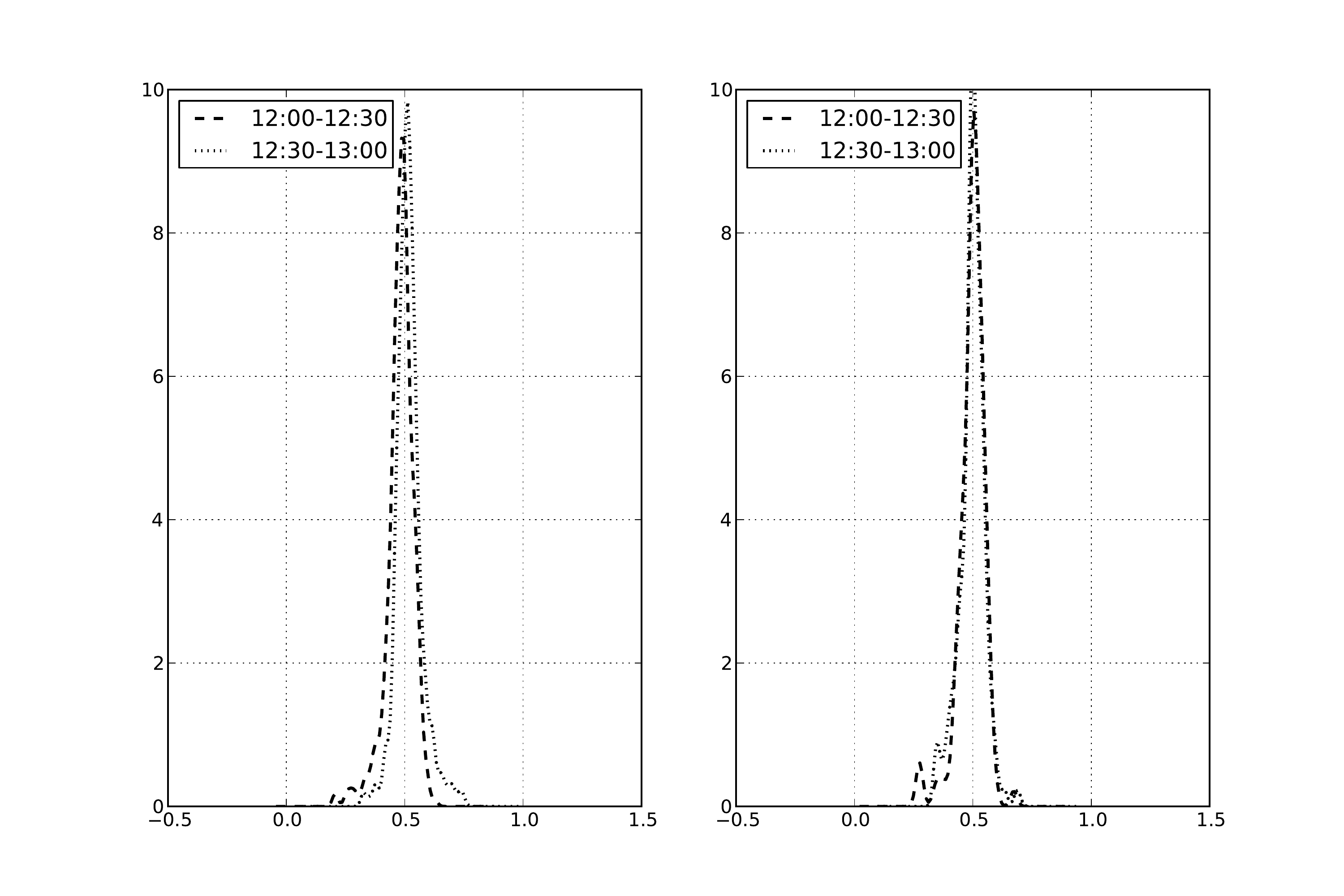}
}
\subfigure[\textbf{Out of
sample}: Estimated densities of the regressed values, K-S distance:
$0.07$, correct classification: $48\%$]{
\includegraphics[trim = 154mm 5mm 10mm 15mm, clip,width =
0.47\textwidth]{plotCDF_timeBuckets_120vs1230_2011_2012_2013_PDF}
}
\subfigure[\textbf{ROC curve.} Area under ROC -- learning set: 0.67, out of
sample: 0.49 ]{
\includegraphics[trim = 10mm 5mm 10mm 10mm, clip,width =
0.8\textwidth]{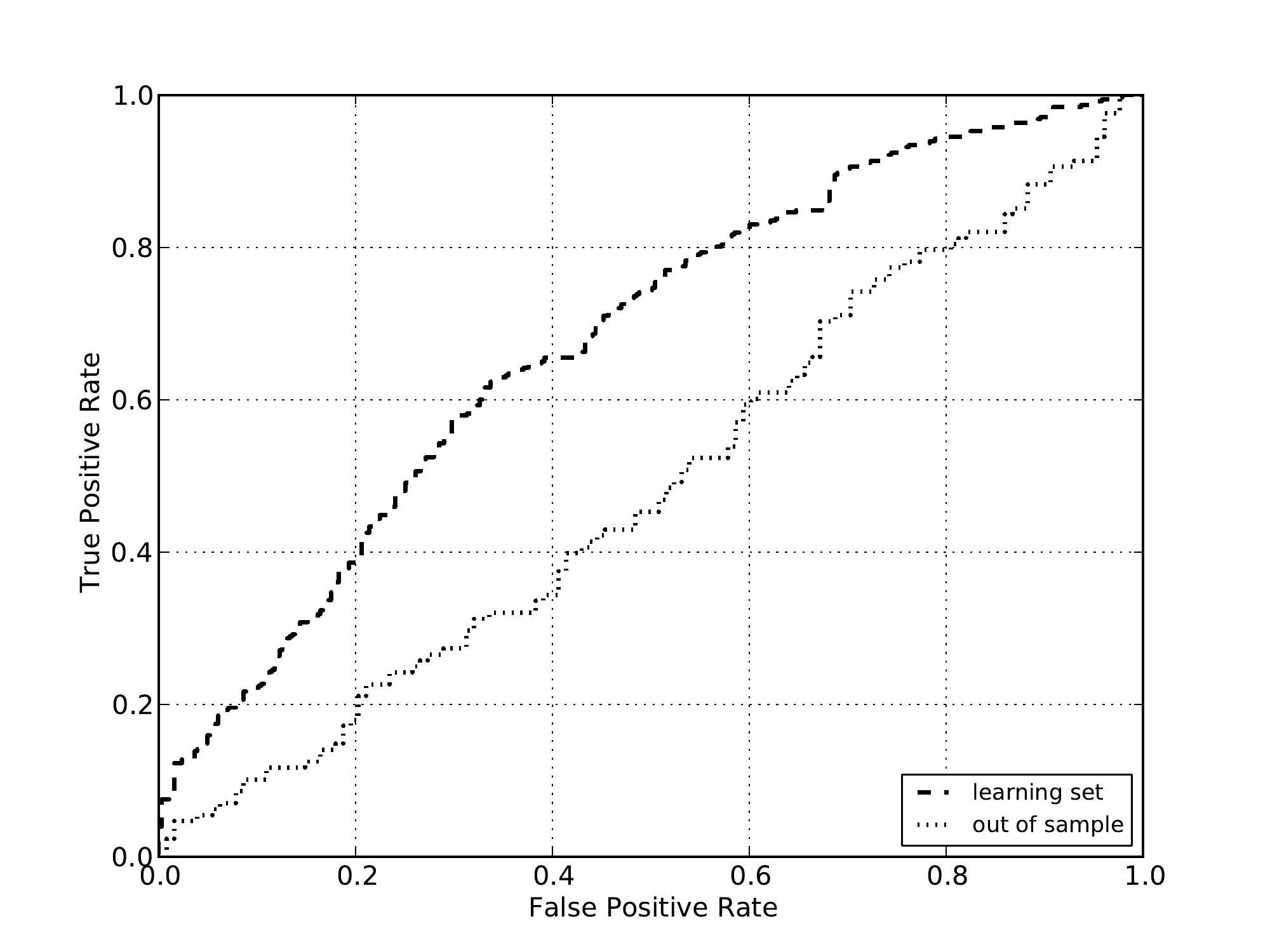}
}
\caption{12:00-12:30 EST versus 12:30-13:00 EST}
\label{fig:120vs1230}
\end{figure}

\paragraph{12:00-12:30 versus 12:30-13:00} Naturally, the market shows
similar behaviour in certain pairs of time intervals. For instance,
the signatures of the data streams observed in the 12:00-12:30 EST and
12:30-13:00 EST intervals indicate no significant difference. Figure
\ref{fig:120vs1230} demonstrates the low-accuracy classification
results in this particular case. 

\subsection{Characterising the traces of trade execution algorithms}\label{sec:AvsB}
The second numerical experiment aims to characterise the traces
of trade execution algorithms. For this purpose, we use data streams
sampled from the FTSE 100 index futures market (NYSE Liffe), which we transform as
described in section \ref{sec:inputData}. The start and finish time of
the sampled streams are determined by the start and completion time of
parent orders generated by two different trade execution algorithms. 
We label these trade execution algorithms by \emph{A} and \emph{B}. 
In the observed time period, the execution team
randomly selected a trade execution algorithm for each trading
signal. There is no significant difference in distributions of order size and start time of parent orders  
that have been executed with algorithm A and algorithm B
respectively.

Note that the time variable is
normalised in each data stream, hence the length of the execution period of the parent
orders is not visible by the signatures.   
Besides the start and completion time of the
parent orders, we have also been given the size of the parent order
(in terms of number of lots) and the direction of the trade (sell or
buy). We did not use any other specifics of the parent orders; in
particular no information on the order submission, order
cancellations and realised trades were used. 

Investors in general aim to minimise their losses when trading. In
particular, they aim to trade at favourable prices and minimise
their market impact. This objective is particularly crucial for market
participants who trade in large volumes. In the light of these facts,
we approached this numerical experiment with moderately high
expectations. Moreover, many trading algorithms operate simultaneously
on the market; some of these algorithms might possess similar
characteristics to the algorithms A and B 
in the focus of our experiment. This can results in noisy observations.

\paragraph{Input streams} The data set contains $453$ and $346$ data
streams that are subsampled from buy parent orders and sell parent orders
respectively over the time period 1/4/2012-1/10/2012. 
 The experiment is focused on parent orders of a trade size
between $5$ and $100$ contracts.

\begin{figure}[ht] 
\centering
\subfigure[\textbf{Learning
set}: Estimated densities of the regressed values, K-S distance: $0.660$,
correct classification: $82.8\%$]{
\includegraphics[trim = 10mm 5mm 154mm 15mm, clip,width =
0.47\textwidth]{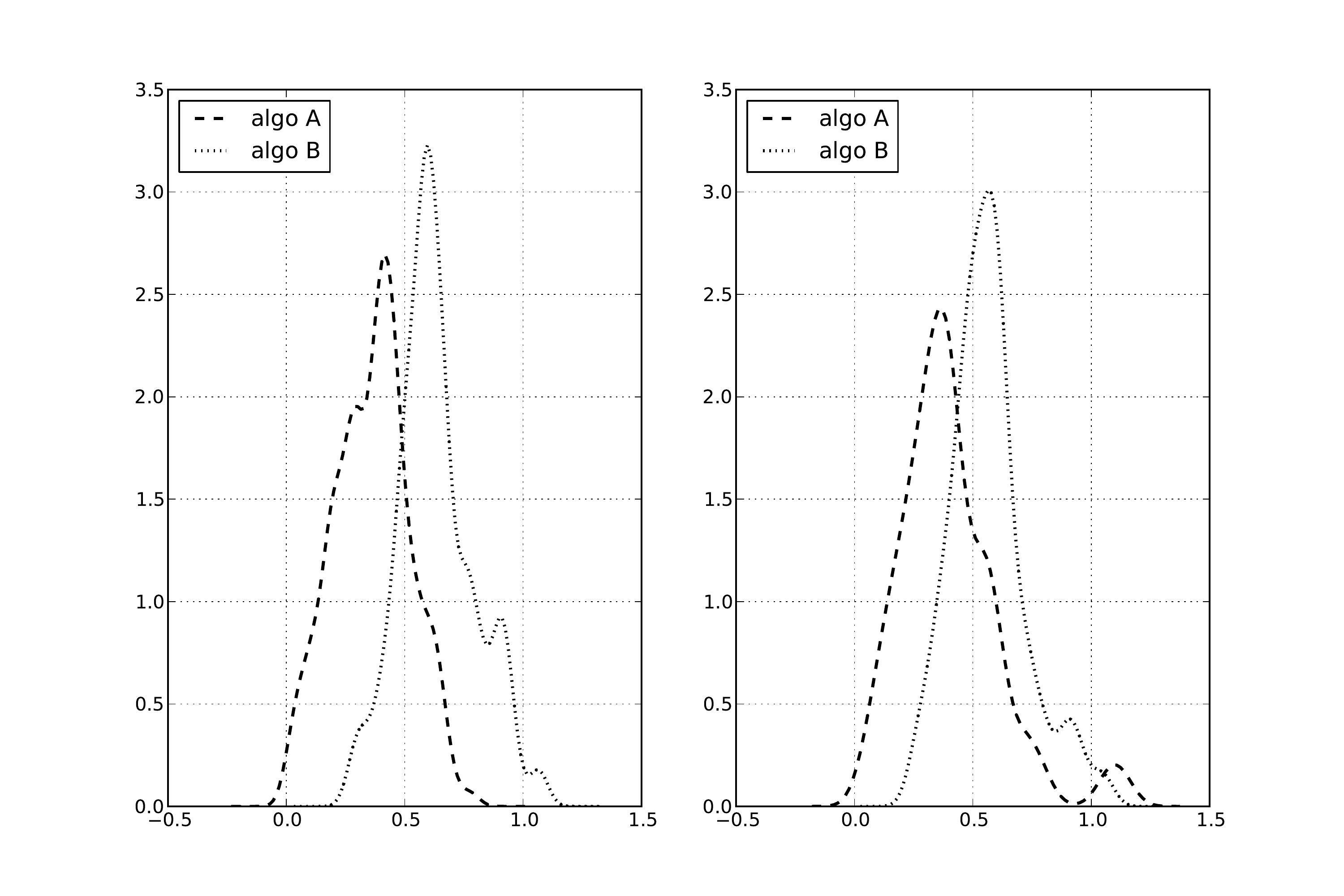}
}
\subfigure[\textbf{Out of
sample}: Estimated densities of the regressed values, K-S distance:
$0.518$, correct classification: $74.3\%$]{
\includegraphics[trim = 154mm 5mm 10mm 15mm, clip,width =
0.47\textwidth]{FEB_results_all_4_var2_CDF_1_PDF}
}
\subfigure[\textbf{ROC curve.} Area under ROC -- learning set: $0.892$, out of
sample: $0.777$ ]{
\includegraphics[trim = 10mm 3mm 10mm 10mm, clip,width =
0.8\textwidth]{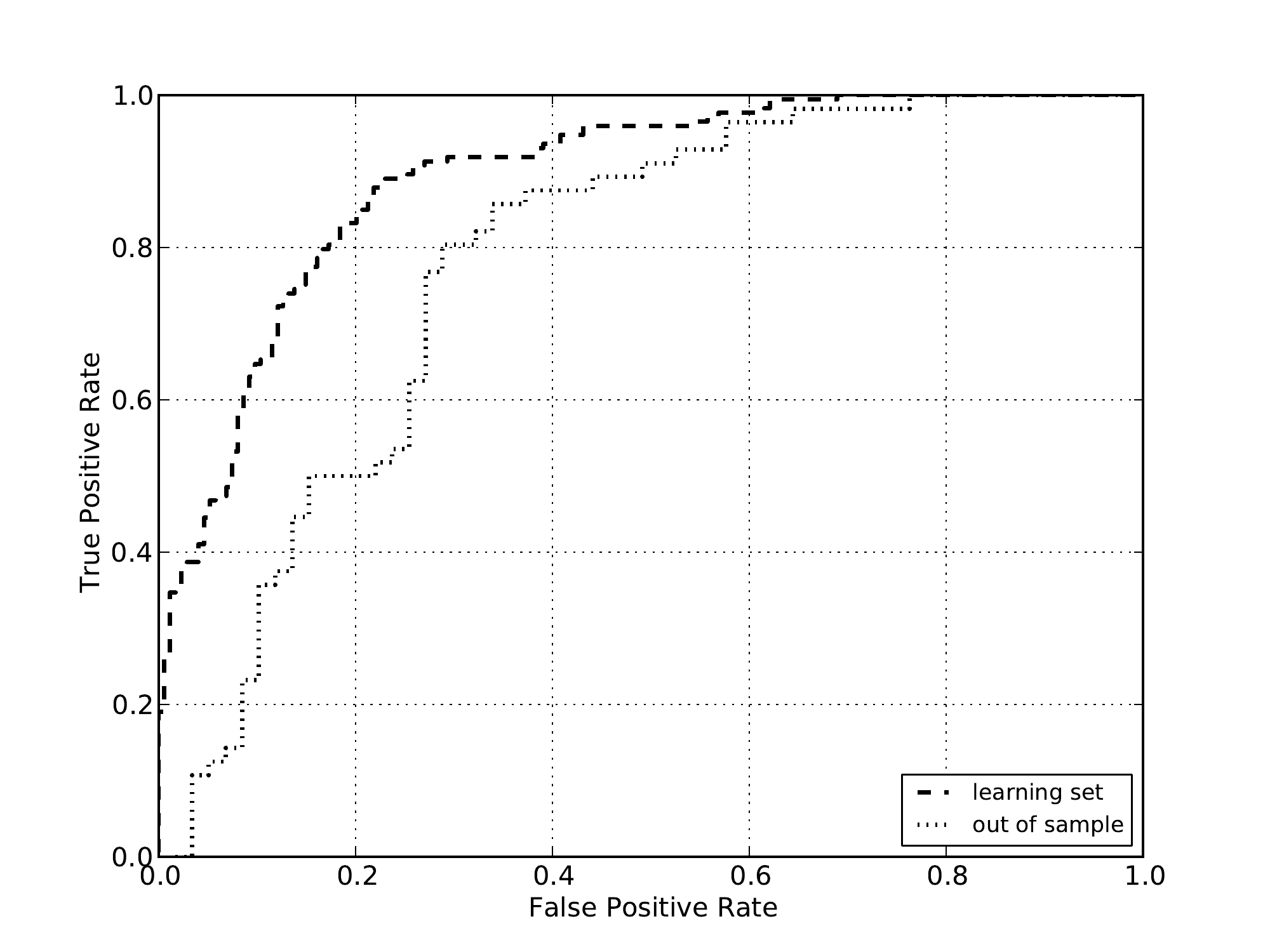}
}
\caption{Algo A versus Algo B on buy orders (1/4/2012-1/10/2012)}
\label{fig:AvsB}
\end{figure}

\paragraph{Classification results}
We selected 75\% of buy (sell) orders randomly 50 times, and for each
learning set we trained
the linear regression based classifier. In each case, the remaining
25\% of the data was used for out of sample testing. In each case we
computed the statistical indicators as described in section \ref{sec:statInd}.
The percentiles of the
computed statistical indicators are shown in table \ref{tab:41buy}
(buy orders) and in table \ref{tab:41sell} (sell orders). 
As a sanity check, we shuffled the algo A and algo B labels randomly
50 times, and for each case we ran the classification and computed the
statistical indicators of the. The 95\%
percentile of these indicators is included in the ``Ref'' columns of
the tables \ref{tab:41buy}
and \ref{tab:41sell} as a reference.

The tables indicate that the signature based learning identifies significant
difference between the trading algorithms A and B. Moreover, it is
more likely to 
perform better on buy orders than on sell orders. 

For a
particular choice of learning set of buy orders, the
distribution of score values and the ROC curve are plotted in figure
\ref{fig:AvsB}. In this particular training, 
the LASSO
shrinkage method identified 56 relevant variables. 
The multi-indices of the top 15 relevant
variables and the corresponding regression coefficients are listed in tables
\ref{table:coeffs_algo_buy}; 
 the indices $1$, $2$, $3$, $4$, $5$ and $6$ correspond to the variables
$u^{\text{lead}}$, $p^{\text{lead}}$,
$s^{\text{lead}}$, $d^{\text{lead}}$, $c^{\text{lead}}$
and $p^{\text{lag}}$ respectively. 

\begin{table}
\centering
\begin{tabular}{|c||c|c|c|c|c|c|c|c|c||c|}
\hline
 & Min & 5\% & 10\% & 25 \% & 50\% & 75 \% & 90\% & 95\% & Max & Ref \\
\hline 
\hline
\multicolumn{11}{|l|}{Kolmogorov-Smirnov distance} \\
\hline
 LS &     0.270 &      0.542 &      0.556 &      0.624 &      0.670 &
 0.707 &      0.746 &      0.763 &      0.802  & \\
 OS &    0.179 &      0.220 &      0.252 &      0.280 &      0.353 &
 0.422 &      0.456 &      0.486 &      0.518  & 0.263\\
\hline
\hline
\multicolumn{11}{|l|}{Area under ROC curve} \\
\hline
  LS &   0.663 &      0.839 &      0.851 &      0.882 &      0.906 &
  0.927 &      0.938 &      0.941 &      0.957 &  \\
   OS &   0.575 &      0.604 &      0.618 &      0.646 &      0.685 &
   0.721 &      0.747 &      0.779 &      0.796  & 0.588\\
\hline
\hline
\multicolumn{11}{|l|}{Ratio of correct classification} \\
\hline
   LS &  0.634 &      0.769 &      0.776 &      0.810 &      0.833 &
   0.852 &      0.871 &      0.879 &      0.899 &  \\
    OS &  0.525 &      0.570 &      0.578 &      0.601 &      0.644 &
    0.670 &      0.703 &      0.709 &      0.744  & 0.572\\
\hline
\end{tabular}
\caption{Percentiles of the classification indicators -- trade
  execution algorithms, buy orders. 
``LS'': learning set, ``OS'': out-of-sample set, ``Ref'': 95\%
percentile of the classification based on randomised labels.}\label{tab:41buy}
\end{table}

\begin{table}
\centering
\begin{tabular}{|c||c|c|c|c|c|c|c|c|c||c|}
\hline
 & Min & 5\% & 10\% & 25 \% & 50\% & 75 \% & 90\% &95\% &  Max & Ref\\
\hline 
\hline
\multicolumn{11}{|l|}{Kolmogorov-Smirnov distance} \\
\hline
 LS &     0.290 &      0.404 &      0.463 &      0.535 &      0.595 &
 0.712 &      0.782 &      0.809 &      0.926 &  \\
 OS &     0.121 &      0.154 &      0.158 &      0.189 &      0.260 &
 0.314 &      0.345 &      0.403 &      0.457  & 0.226\\
\hline
\hline
\multicolumn{11}{|l|}{Area under ROC curve} \\
\hline
  LS &  0.677 &      0.744 &      0.780 &      0.830 &      0.871 &
  0.924 &      0.955 &      0.966 &      0.994  & \\
  OS &   0.525 &      0.538 &      0.544 &      0.578 &      0.609 &
  0.650 &      0.680 &      0.699 &      0.715  & 0.579\\
\hline
\hline
\multicolumn{11}{|l|}{Ratio of correct classification} \\
\hline
   LS & 0.644 &      0.701 &      0.730 &      0.766 &      0.795 &
   0.853 &      0.888 &      0.902 &      0.960  & \\
 OS &   0.473 &      0.519 &      0.529 &      0.551 &      0.584 &
 0.618 &      0.639 &      0.660 &      0.685  & 0.568\\
\hline
\end{tabular}
\caption{Percentiles of the classification indicators -- trade
  execution algorithms, sell orders.
``LS'': learning set, ``OS'': out-of-sample set, ``Ref'': 95\%
percentile of the classification based on randomised labels.}\label{tab:41sell}
\end{table}


\begin{table}[ht]
\centering
\begin{tabular}{|c|r||c|r||c|r|}
\hline
Multi-index & \multicolumn{1}{c||}{coefficient} & Multi-index & \multicolumn{1}{c|}{coefficient} & Multi-index & \multicolumn{1}{c|}{coefficient} \\
\hline
(1,6,4,1) &   0.070366 & (1,1,6,4) &   0.061548 & (1,4,2,3) &  -0.047923 \\
(1,4,2,5) &   0.043171 & (1,4,4,1) &  -0.042334 & (2,3,5,4) &  -0.040247 \\
(4,1,3,4) &  -0.039719 & (6,3,6,5) &   0.037544 & (4,2,4,2) &   0.032859 \\
(3,4,4,3) &  -0.031906 & (2,1,2,5) &   0.031789 & (3,1,6,1) &  -0.026957 \\
(4,2,4) &   0.026546 & (2,4,2,3) &  -0.025268 & (3,2,6,4) &  -0.025219 \\
\hline
\end{tabular}
\caption{Classification of trade execution algorithms on buy orders \\
  Estimated regression coefficients}
\label{table:coeffs_algo_buy}
\end{table}


\subsection{Robustness of  classification}\label{sec:localAvsB}



\begin{figure}[ht] 
\centering
\subfigure[Ratio of correct classification]{
\includegraphics[trim = 10mm 5mm 10mm 10mm, clip, width =
0.48\textwidth]{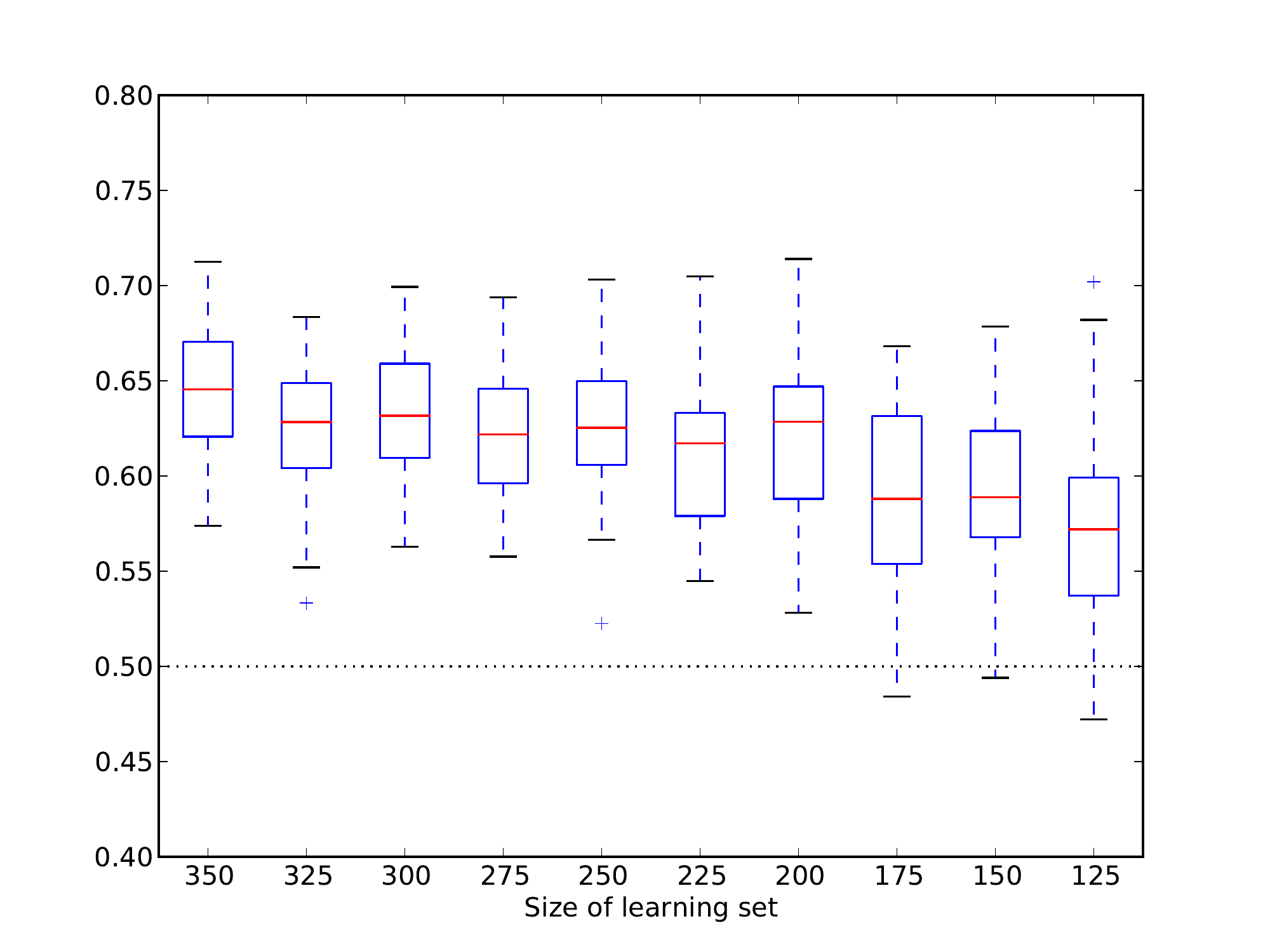}
}
\subfigure[Area under ROC curve]{
\includegraphics[trim = 10mm 5mm 10mm 10mm, clip, width =
0.48\textwidth]{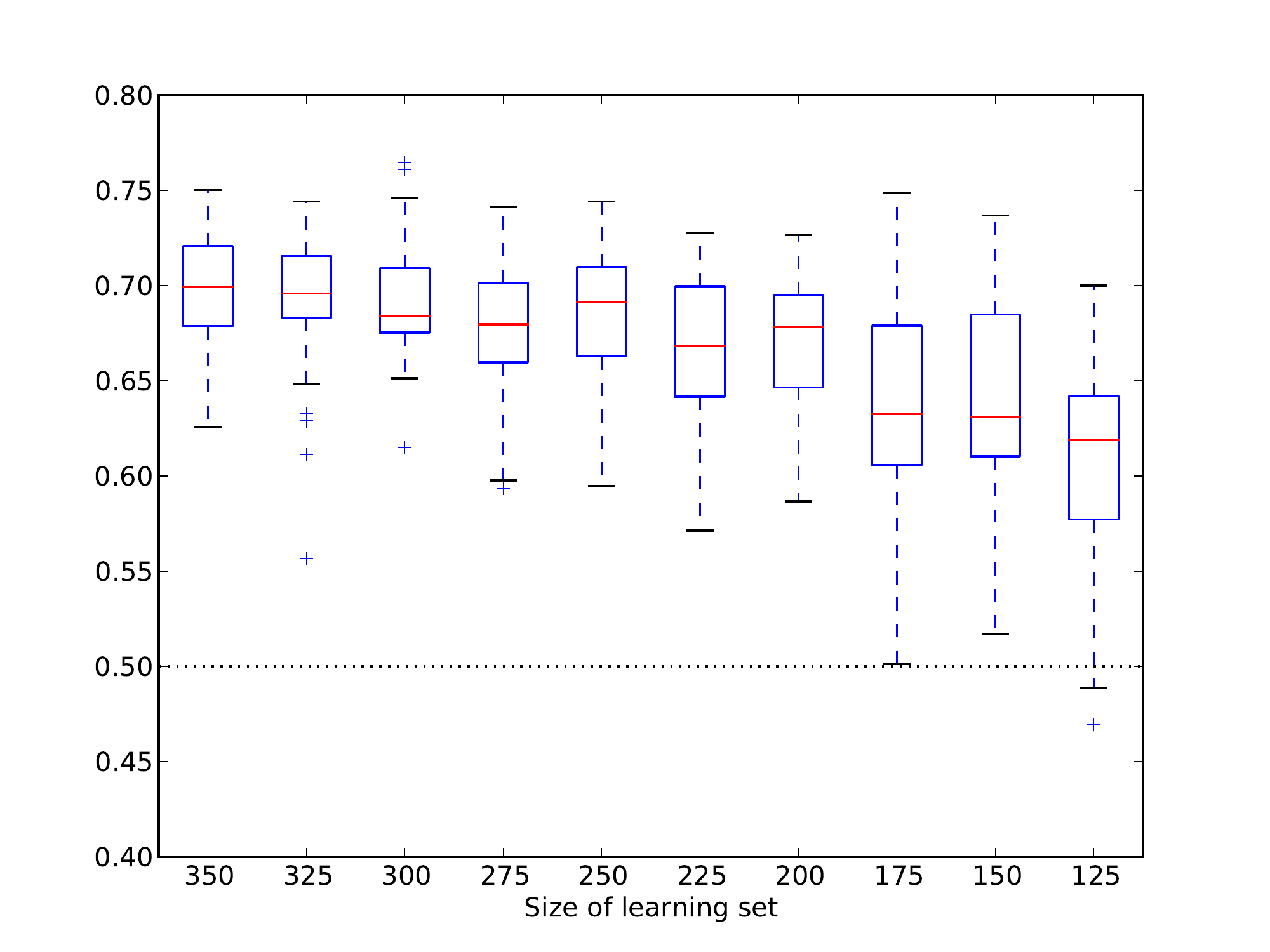}
}
\subfigure[Kolmogorov-Smirnov distance]{
\includegraphics[trim = 10mm 5mm 10mm 10mm, clip, width =
0.48\textwidth]{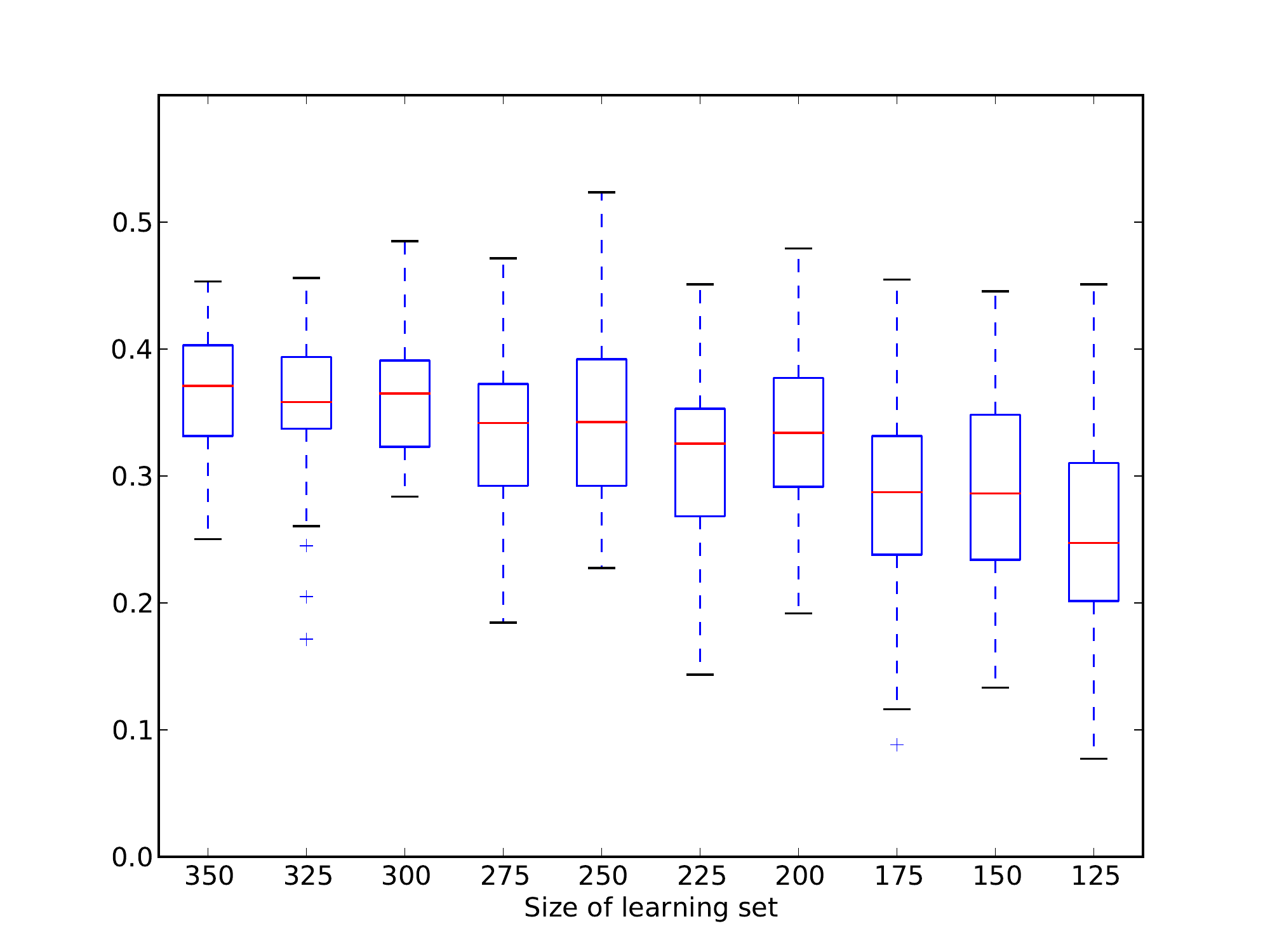}
}
\caption{Algo A vs Algo B -- Randomised learning sets. The indicators describe the performance on a fixed out of
  sample test.}
\label{fig:AvsBLocalised1}
\end{figure}

In order to get a better understanding of the prediction power of our
classification method, we trained the tool on randomised learning sets
of different sizes. In particular, the learning sets were randomly
sampled from the first 353 data streams that corresponded to buy
orders 50 times for each learning set size. Moreover, 
each trained classifier was tested on the same out of sample test. 
Figure \ref{fig:AvsBLocalised1} shows the distribution of the
various statistical indicators corresponding to trainings as a
function of the size of the learning set. We observe that the
classification performance is reasonable robust when the learning set
has 200 streams or more. For learning set of less than 200 streams the
variance of the indicators increases, and the classification accuracy
drops.



\section{Conclusions}
We use the terms in the signature of a data stream to define features
for supervised learning algorithms. In particular, we  extract information
from noisy market data and classify data streams along various
properties with a degree of high
accuracy. The numerical examples demonstrate that the signature-based
classification has great potential in machine learning.

\newpage
\bibliographystyle{amsplain}
\bibliography{bibFile}

\providecommand{\bysame}{\leavevmode\hbox to3em{\hrulefill}\thinspace}
\providecommand{\MR}{\relax\ifhmode\unskip\space\fi MR }
\providecommand{\MRhref}[2]{%
  \href{http://www.ams.org/mathscinet-getitem?mr=#1}{#2}
}
\providecommand{\href}[2]{#2}
\begin{thebibliography}{10}

\bibitem{lasso}
Peter B{\"u}hlmann and Sara van~der Geer, \emph{{Statistics for
  High-Dimensional Data: Methods, Theory and Applications}}, Springer Series in
  Statistics, Springer, 2011.

\bibitem{Chen57}
Kuo-Tsai Chen, \emph{{Integration of paths, geometric invariant and generalized
  Campbell-Baker-Hausdorff formula}}, Ann. Math. \textbf{65} (1957), no.~1,
  163--178.

\bibitem{Chen58}
\bysame, \emph{{Integration of paths -- a faithful representation of paths by
  non-commutative formal power series}}, Trans. Amer. Math. Soc. \textbf{89}
  (1958), 395--407.

\bibitem{FrizVictoir}
Peter Friz and Nicolas Victoir, \emph{{Multidimensional Stochastic Processes as
  Rough Path}}, Cambridge University Press, 2010.

\bibitem{gyurkoThesis}
Lajos~Gergely Gyurk\'o, \emph{{Numerical Methods for Approximating Solutions to
  Rough Differential Equations}}, {DPhil Thesis}, University of Oxford, 2009.

\bibitem{RPNumSDE}
Lajos~Gergely Gyurk\'o and Terry Lyons, \emph{{Rough Paths based Numerical
  Algorithms in Computational Finance}}, Mathematics in Finance
  (Santiago~Carrillo Men\'endez and Jos\'e~Luis Fern\'andez~P\'erez, eds.),
  American Mathematical Society, Real Sociedad Matem\'atica Espa\~{n}ola, 2010,
  pp.~397--405.

\bibitem{uniqueSig}
Ben Hambly and Terry Lyons, \emph{{Uniqueness for the signature of a path of
  bounded variation and the reduced path group}}, Annals of Mathematics
  \textbf{171} (2010), no.~1, 109--167.

\bibitem{KloedenPlaten}
E.~Peter Kloeden and Eckhard Platen, \emph{{Numerical Solution of Stochastic
  Differential Equations}}, Springer-Verlag Berlin Heidelberg New York, 1999.

\bibitem{Kusuoka04}
Shigeo Kusuoka, \emph{{Approximation of expectation of diffusion process based
  on Lie algebra and Malliavin calculus}}, Adv. Math. Econ. \textbf{6} (2004),
  69--83.

\bibitem{roughPaths98}
Terry Lyons, \emph{{Differential equations driven by rough signal}}, Revista
  Matematica Iberoamericana \textbf{14} (1998), no.~2, 215--310.

\bibitem{SaintFlour}
Terry Lyons, Michael Caruana, and Thierry L\'evy, \emph{{Differential Equations
  Driven by Rough Paths}}, Ecole d'Et\'e de Porbabilit\'es de Saint-Flour XXXIV
  - 2004, Lecture Notes in Mathematics, Springer, 2007.

\bibitem{NiLyons}
Terry Lyons, Hao Ni, and Daniel Levin, \emph{{Learning from the past,
  predicting the statistics of the future, learning an evolving system}},
  {Preprint}, 2013.

\bibitem{roughPaths02}
Terry Lyons and Zhongmin Qian, \emph{{System Control and Rough Paths}}, Oxford
  Mathematical Monographs, Clarendon Press, 2002.

\bibitem{cubature}
Terry Lyons and Nicolas Victoir, \emph{{Cubature on Wiener Space}}, Proc. R.
  Soc. Lond. \textbf{A 460} (2004), 169--198.

\end{thebibliography}
\end{document}